\documentclass[10pt, conference]{IEEEtran}

\usepackage{cite}
\usepackage{amsmath,amssymb,amsfonts}
\usepackage{algorithmic}
\usepackage{textcomp}
\def\BibTeX{{\rm B\kern-.05em{\sc i\kern-.025em b}\kern-.08em
    T\kern-.1667em\lower.7ex\hbox{E}\kern-.125emX}}

\usepackage{hyperref}
\usepackage{color}
\usepackage{xcolor}
\usepackage{graphicx}
\usepackage{listings}
\usepackage{multirow}
\usepackage{fancyhdr}
\usepackage{xspace}
\usepackage{times}
\usepackage{tikz}
\usepackage{wrapfig}
\usepackage[font=small]{caption}

\graphicspath{ {./images/} }

\lstdefinestyle{CUDA}{language=C++,
    commentstyle=\color{blue},
    numberstyle=\scriptsize,
    captionpos=b,
    numbers=left,
    xleftmargin=1em,
    frame=single,
    numbersep=5pt,
    basicstyle=\color{black}\ttfamily\scriptsize,
    breaklines=true
}

\definecolor{codegreen}{rgb}{0,0.6,0}
\definecolor{codegray}{rgb}{0.5,0.5,0.5}
\definecolor{codepurple}{rgb}{0.58,0,0.82}
\definecolor{backcolour}{rgb}{0.95,0.95,0.92}

\lstdefinestyle{DSL}{
    language=C++,
    literate={->}{{$\rightarrow$}}1,
    morekeywords={int,int8,int16,int32,float,double,bool,vector},
    morekeywords={Matrix,for,if,else,Params,CodeGen},
    backgroundcolor=\color{lightlightgray},
    commentstyle=\color{codegreen},
    keywordstyle=\color{magenta},
    numberstyle=\tiny\color{codegray},
    stringstyle=\color{codepurple},
    basicstyle=\ttfamily\footnotesize,
    breakatwhitespace=false,         
    breaklines=true,                 
    captionpos=b,                    
    keepspaces=true,                 
    numbers=left,                    
    numbersep=5pt,                  
    showspaces=false,                
    showstringspaces=false,
    showtabs=false,                  
    tabsize=2 
}

\definecolor{shadecolor}{RGB}{220,220,220}
\newcommand{\mybox}[1]{\par\noindent\colorbox{shadecolor}
{\parbox{\dimexpr\textwidth-2\fboxsep\relax}{#1}}}

\begin{document}


\title{Automatic Kernel Generation for Volta Tensor Cores}

\author{\IEEEauthorblockN{Somashekaracharya G. Bhaskaracharya}
\IEEEauthorblockA{\textit{NVIDIA} \\
sbhaskaracha@nvidia.com}
\and
\IEEEauthorblockN{Julien Demouth}
\IEEEauthorblockA{\textit{NVIDIA} \\
jdemouth@nvidia.com}
\and
\IEEEauthorblockN{Vinod Grover}
\IEEEauthorblockA{\textit{NVIDIA} \\
vgrover@nvidia.com}}

\maketitle

\begin{abstract}


A commonly occurring computation idiom in neural networks is to perform some
pointwise operations on the result of a matrix multiplication. Such a
sequence of operations is typically represented as a computation graph in
deep learning compilers. When compiling to a GPU target, these computations
can be individually mapped to manually tuned implementations provided by
libraries such as cuBLAS and cuDNN. These libraries also provide
off-the-shelf support for targeting tensor cores in NVIDIA GPUs, which can
lead to huge performance boosts through their specialized support for
mixed-precision matrix math. Alternatively, tensor cores can be programmed
directly using CUDA APIs or inline assembly instructions, which opens up the
possibility of generating efficient CUDA kernels automatically for such
computations.

Automatic kernel generation is particularly crucial when it is beneficial to
generate efficient code for an entire computation graph by fusing several
operations into a single device function instead of invoking a separate
kernel for each of them. Polyhedral compilation techniques provide a
systematic approach for the analysis and transformation of a sequence of
affine loop-nests. In this paper, we describe a polyhedral approach to
generate efficient CUDA kernels for matrix multiplication using inline
assembly instructions for programming tensor cores on NVIDIA Volta GPUs.
Furthermore, we build on this approach to generate fused kernels for
computation sequences involving matrix multiplication and pointwise
operations such as bias addition, ReLU activation etc. Experimental
evaluation of these techniques show that automatically generated kernels can
provide significantly better performance than manually tuned library
implementations, with speedups ranging up to 2.55$\times$.


\end{abstract}




\begin{IEEEkeywords}
matmul, polyhedral compilation, tensor cores
\end{IEEEkeywords}

\section{Introduction}

Tensor cores in NVIDIA GPUs are processing cores specialized for matrix
operations. They provide huge boosts in throughput and efficiency by
performing several mixed-precision matrix-multiply and accumulate
calculations in a single operation. Consequently, tensor cores can
significantly speed up generalized matrix-multiply (GEMM) and convolutions
(as implicit GEMMs), both of which are used heavily in deep learning systems
and other computational applications. CUDA libraries such as cuBLAS~\cite{cublas} and
Cutlass~\cite{cutlass} provide off-the-shelf support for leveraging tensor core capabilities
through manually tuned implementations of GEMM for various input matrix
layouts (row-major and column-major). Furthermore, cuDNN~\cite{chetlur2018cudnn} is a widely used
GPU-accelerated library with extensive support for various deep learning
primitives.

Deep learning computations can be modeled as dataflow graphs where each
node represents a specific computation such as matmul, convolution, pointwise
operations etc. A commonly occurring computation idiom in neural networks is
that of a matmul feeding a bias add, which then drives an activation
function such as ReLU. These computations can be specified through various
deep learning frameworks such as TensorFlow~\cite{abadi2016osdi},
PyTorch~\cite{paszke2019nips}. A naive approach to map these computations
onto a GPU would invoke separate kernels for each computation node in an
execution order dictated by the dataflow dependences. However, the repeated
round-trip through global memory is inefficient and can be eliminated if
multiple operations are fused into the same kernel. So, for the
above example, the entire computation should be ideally
mapped to a single kernel that performs matmul as well as its downstream
pointwise operations, without the overhead of writing the intermediate
results to global memory. In this context, it is imperative to explore kernel
generation techniques not just for matmul but also for matmul with such
pointwise operations in its epilogue or prologue i.e., downstream or upstream
to it in the computation graph.

Automatic kernel generation for tensor cores requires direct
programmatic access to tensor cores.
On a Volta GPU, tensor cores can be programmed directly using the
\textit{mma.sync.m8n8k4} PTX instruction~\cite{cudaToolkit} for
half-precision floating point type. The operation is defined for each
quad-pair, i.e, a group of 8 threads. The \textit{mma.sync.m8n8k4}
instruction for half-precision floats (the \textit{fp16} data type) on Volta requires each
thread in a quad pair to own fragments of the input matrix to a matrix
multiply-and-accumulate operation. Each fragment consists of 4
half-precision floats. Consequently, the input fragments from all the threads
in a quad-pair constitute the two 8$\times$4 and 4$\times$8 input matrices
for the matrix-multiply-and-accumulate operation (mma). The resulting
8$\times$8 matrix is distributed across the quad-pair's threads, and each
thread owns a piece of the accumulator matrix called the accumulator
fragment, which is of size 1$\times$8.

Another way to program tensor cores is through the wmma CUDA
APIs~\cite{wmma2017}. Unlike an \textit{mma.sync.m8n8k4} instruction, the
wmma operation is defined for an entire warp. A wmma fragment distribution is
opaque, and is target dependent. On the other hand, the distribution of
\textit{mma.sync} fragments is pre-defined and not target architecture
dependent. Furthermore, as we shall see later, the \textit{m8n8k4} mma operation can
be composed into higher-order macro-MMA operations in a number of ways to
take advantage of wider loads and stores at every level of the memory
hierarchy. Due to this low-level control, the \textit{mma.sync.m8n8k4} instruction lends itself well to implement efficient kernels for deep learning graphs, with
ample scope for kernel fusion.

Our focus is on generating efficient kernels for computation graphs that
involve matmul and pointwise operations such as add,
subtract, and activation functions such as ReLU, Sigmoid and Tanh. In
particular, the scope of this work is restricted to the following computation
idioms.

\begin{itemize}
    \item matmul without any epilogue or prologue
    \item matmul with pointwise operations in its prologue
    \item matmul with pointwise operations in its epilogue
    \item matmuls of the same shape feeding a pointwise operation
\end{itemize}

These operations can be expressed as affine loop-nests. Polyhedral frameworks
have proven to be effective for analyzing and transforming a sequence of
affine loop-nests~\cite{uday08pldi, verdoolaege2013Taco, baskaran08ics}.
Several deep learning compiler frameworks~\cite{vasilache2018tcCoRR,
mlir2019, schreiber2020impact} support polyhedral compilation techniques in
conjunction with other analysis tools and intermediate representations. In
this paper, we describe a polyhedral approach to automatically generate
efficient kernels for Volta tensor cores given a high-level computation DAG
where each node represents a matmul or a pointwise operation such as add,
subtract, ReLU, Sigmoid, Tanh. To summarize, our contributions are as
follows.

\begin{itemize}

    \item We describe polyhedral techniques to automatically generate
    efficient kernels that implement matmul using the \textit{mma.m8n8k4} PTX 
    instruction on a Volta GPU, by composing the \textit{m8n8k4} mma tile into
    higher-order macro-MMA operations of shapes \textit{m16n16k8} and
    \textit{m32n32k8} for realizing a warp-level MMA.

    \item We then build on the above approach to automatically perform kernel
    fusion for some commonly occurring computation idioms involving matmul
    and pointwise operations.

    \item We implement and evaluate our approach for these computation idioms
    on various problem sizes and demonstrate significant speedups over
    manually-tuned implementations provided by standard libraries such as
    cuBLAS and cuDNN.

\end{itemize}

Section~\ref{sec:background} provides the necessary background and introduces
the notation use in later sections. Sections~\ref{sec:core1}, 
~\ref{sec:core2} and ~\ref{sec:code-gen} describe the compute decomposition
and data movement required for mapping a matmul computation to Volta tensor
cores. Kernel fusion using similar decompositions is discussed in
Section~\ref{sec:core3}. Experimental evaluation of these techniques is
provided in Section~\ref{sec:experimental}. Related work and conclusions are
presented in Sections~\ref{sec:related-work} and ~\ref{sec:conclusions}
respectively.

\section{Background}
\label{sec:background}

This section provides the notation and background for the techniques we
present in the rest of this paper.


\colorlet{MyBlue}{blue!50}
\newcommand{\lightercolor}[3]{
    \colorlet{#3}{#1!#2!white}
}
\lightercolor{MyBlue}{50}{MyBlueLight}

\subsection{Programming Tensor Cores}

The GPU compute hierarchy consists of blocks of threads which are
organized into warps. Each warp consists of 32 threads. These
threads can be further partitioned into 8 \textit{quads}, i.e., threads
$0-3$, $4-7$, $8-11$, $12-15$, $16-19$, $20-23$, $24-27$, $28-31$
respectively. A Volta \textit{mma.sync.m8n8k4} instruction is executed by a
pair of quads. For example, threads $0-3$ and $16-19$ constitute quad-pair
$QP_0$; threads $4-7$, $20-23$ constitute quad-pair $QP_1$ and so on.

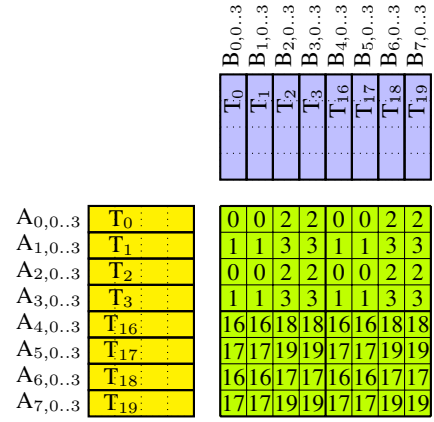
\begin{figure}[t!]{}
    \centering
    \begin{tikzpicture}[scale=0.35]
        \filldraw [fill=yellow, draw=yellow] (-5, 8) rectangle (-1, 0);
        \draw [dotted] (-4 - 1, 8) grid (-1, 0);
        \foreach \count in {0, 1, ..., 7}
        {
            \draw [thick](-5, 8 - \count) rectangle (-1, 8 - \count - 1);
            \node at (-6.5, 8 - \count - 0.5) {\small {A$_{\count, 0..3}$} \normalsize};
            \pgfmathsetmacro{\threadId}{int(int(\count / 4) * 12 + \count)}
            \node at (-3.75, 8 - \count - 0.5) {\small {T$_{\threadId}$} \normalsize};
        }

        \filldraw [fill=MyBlueLight, draw=MyBlueLight] (0, 9) rectangle (8, 13);
        \draw [dotted] (0, 8 + 1) grid (8, 12 + 1);
        \foreach \x in {0, 1, ..., 7}
        {
            \draw [thick] (\x, 9) rectangle (\x + 1, 13);
            \node [rotate=90] at (\x + 0.5, 14.5) {\small B$_{\x, 0..3}$ \normalsize};
            \pgfmathsetmacro{\threadId}{int(int(\x / 4) * 12 + \x)}
            \node [rotate=90] at (\x + 0.5, 12) {\small {T$_{\threadId}$} \normalsize};
        }

        \newcounter{column}
        \newcommand\setrow[9]{
            \foreach \ycor in {#9}
            {
                \setcounter{column}{0.5}
                \foreach \num in {#1, #2, #3, #4, #5, #6, #7, #8} {
                    \edef\xcor{\value{column} + 0.5}
                    \node[anchor=center] at (\xcor, \ycor) {\small \num \normalsize};
                    \stepcounter{column}
                }
            }
        }

        \filldraw [fill=lime, draw=lime] (0, 0) rectangle (8, 8);
        \draw [] (0, 0) grid (8, 8);
        \draw[thick, xscale=4, yscale=4] (0, 0) grid (2, 2);

        \setrow {0}{0}{2}{2}{0}{0}{2}{2}{7.5}
        \setrow {1}{1}{3}{3}{1}{1}{3}{3}{6.5}
        \setrow {0}{0}{2}{2}{0}{0}{2}{2}{5.5}
        \setrow {1}{1}{3}{3}{1}{1}{3}{3}{4.5}
        \setrow {16}{16}{18}{18}{16}{16}{18}{18}{3.5}
        \setrow {17}{17}{19}{19}{17}{17}{19}{19}{2.5}
        \setrow {16}{16}{17}{17}{16}{16}{17}{17}{1.5}
        \setrow {17}{17}{19}{19}{17}{17}{19}{19}{0.5}
    \end{tikzpicture}
\caption{\small The shape and data distribution for an
\texttt{mma.sync.m8n8k4.row.col} operation. Matrix A is row major while
matrix B is column major. \normalsize} 
\label{fig:mma-m8n8k4-row-col}
\vspace{-0.5cm}
\end{figure}

As shown in Figure~\ref{fig:mma-m8n8k4-row-col}, each mma operation is of
shape \textit{m8n8k4}, i.e., input matrices of shape 8$\times$4 and 4$\times$8 are
multiplied and added to an accumulator matrix of shape 8$\times$8. The
elements in the input matrices are half-precision floats (\textit{fp16} type)
whereas the accumulator matrix elements are of type \textit{fp32} (a version
of \textit{mma.m8n8k4} with \textit{fp16} accumulators also exists on Volta,
although we consider the \textit{fp32} version in this paper). The input
matrices can be in either row or column major format with different
specializations of the instruction provided for different layout
combinations. Each thread in a quad pair owns an input data
\textit{fragment}, which is nothing but 4 half-precision floating point
values. As shown in Figure~\ref{fig:mma-m8n8k4-row-col}, thread $0$ owns 4
contiguous \textit{fp16} values from the first row and column of the input
matrices A and B respectively; thread $1$ owns the second row and column and
so on. Furthermore, each thread owns an \textit{accumulator fragment}, which
consists of $8$ elements from the accumulator matrix. In
Figure~\ref{fig:mma-m8n8k4-row-col}, the thread indices specified in the
cells of the 8$\times$8 accumulator matrix indicate the distribution of the
data elements among the different threads.

Finally, with each quad-pair performing a single mma operation, an entire
warp can perform 4 mma operations, each of shape \textit{m8n8k4}. More details on the
\textit{mma.sync.m8n8k4} instruction in Volta can be found in the CUDA
toolkit documentation~\cite{cudaToolkit}.

\subsection{Polyhedral Model}

Polyhedral model is a mathematical representation of affine
loop-nests, where the loop bounds and array access expressions are affine
combinations of enclosing loop iterators and program parameters. In the
polyhedral representation each execution instance of a
statement $S$ is represented as an integer point within a polyhedron. The
faces of the polyhedron correspond to the bounds on the enclosing loops and
the dimensionality of the polyhedron is nothing but the number of enclosing
loops. The integer points within the polyhedron capture the
iteration domain $I_S$ of the statement. Each statement may access arrays
whose dataspaces can also be similarly defined as polyhedra. Consequently, an
access relation $I_S \rightarrow A$ mapping an iteration domain $I_S$ to a
dataspace $A$ can be used to specify accesses to array $A$ performed by a
statement $S$. Similarly, relations between iteration spaces represent RAW,
WAR and WAW dependences. Furthermore, the execution schedule of the statement
instances is specified by mapping the execution instances to
multi-dimensional logical timestamps, whose lexicographic ordering gives the
execution order.

The Integer Set Library (ISL)~\cite{isl} can be used to represent and
manipulate such a polyhedral representation. Given the polyhedral
representation of a loop-nest or a sequence of loop-nests, the polyhedral
scheduler in ISL can be used to determine a valid schedule, such that all
dependences are satisfied. ISL also provides facility for manipulating a
schedule through its schedule tree representation~\cite{verdoolaege14IMPACT}.

\section{Problem Statement}

Our focus is on automatic kernel generation for computation DAGs where the
nodes represent matmul or pointwise operations. Since each of these
operations can be specified as affine loop-nests, each node also encapsulates
its corresponding polyhedral representation -- the iteration space, data
space, access relations as well as the dependence relations. Such a
computation DAG serves as the input to our kernel generation problem. DSLs
such as Halide~\cite{ragan-kelley13pldi}, Tensor
Comprehensions~\cite{vasilache2018tcCoRR} can be used to derive such a
DAG.

\footnotesize
\lstset{style=CUDA}
\begin{lstlisting}[mathescape=true, caption={\small Matmul + Bias + ReLU \normalsize},label={lst:matmul-bias-relu}]
for(i = 0; i $<$ M; ++i)
  for(j = 0; j $<$ N; ++j)
    for(k = 0; k $<$ K; ++k)
/*S1*/ C[i, j] = mul_acc(C[i, j], A[i, k], B[k, j]);

for(i = 0; i $<$ M; ++i)
  for(j = 0; j $<$ N; ++j)
/*S2*/ E[i, j] = relu_add(C[i, j], bias[i, j]);
\end{lstlisting}
\normalsize

Listing~\ref{lst:matmul-bias-relu} shows an example where the matmul result
is fed to pointwise operations -- bias add followed by the ReLU
activation function. Each loop-nest maps to a separate node in the
computation DAG. Now, the polyhedral model would consist of the
following.

\begin{itemize}
    \item The iteration domains, $I_1$ and $I_2$, for the statements $S_1$
    and $S_2$.
    
    \item Dataspaces, $M_C$ and $M_E$, written to by $S_1$ and
    $S_2$ respectively.
    
    \item Write access relations, $I_1 \rightarrow M_C$ and $I_2 \rightarrow
    M_E$, specifying write accesses performed on $M_C$ and $M_E$
    respectively.
    
    \item Read access relations, $\{I_1
    \rightarrow M_C, I_1 \rightarrow M_A, I_1 \rightarrow M_B, I_2
    \rightarrow M_{bias}, I_2 \rightarrow M_C\}$, specifying the read accesses performed on the
    dataspaces $M_C, M_A$, $M_B$ and $M_{bias}$ respectively.
    
    \item RAW dependences, which consist of both intra-node and inter-node
    dependences.
\end{itemize}

Additionally, to facilitate code generation, the following attributes about
each statement are tagged on to its iteration space.

\begin{itemize}
    \item \textit{Expression Tree.} The operations that are to be performed
    as part of the statement are encoded in the form of an expression tree.
    Each internal node in the expression tree corresponds to an operation.
    The result of the root operation is written to the output dataspace. In
    Listing~\ref{lst:matmul-bias-relu}, statement $S_2$ performs a bias add
    and ReLU activation, represented by the compound operation
    \texttt{relu\_add}. Its expression tree consists of two operations --
    ReLU and add, with ReLU as the root operation.

    \item \textit{Write and read access relations}: These are the leaves of
    the expression tree and are used to determine the data
    accesses performed by a statement instance, thereby providing
    the operands for the parent operation nodes in the tree.
\end{itemize}

In the following sections, we first discuss the problem of generating
efficient Volta kernels for matmul and then build on this approach to
generate fused kernels for longer computation sequences.

\section{Matmul Compute Decomposition}
\label{sec:core1}

The ISL scheduler gives an outer-parallel schedule, which
is a good starting point for mapping the computation to a GPU.
The matmul loop in Listing~\ref{lst:matmul-bias-relu} has such a schedule. We
now describe the compute decomposition for mapping this 3-d loop nest to the
GPU compute hierarchy of blocks, warps and threads in order to target tensor
cores.

\subsection{Macro-MMA}

MMA operations with the shape \textit{m8n8k4} are defined for a quad-pair. But they
can be composed into higher-order macro-MMA operations that are performed by
an entire warp. Two such macro-MMA compositions that we employ are of shapes
16$\times$16$\times$8 and 32$\times$32$\times$8.

\begin{figure}[t!]
    \centering{\begin{tikzpicture}[scale=0.26]
\small

    \draw [dotted] (0, 16 + 0.5) grid (16, 24 + 0.5);
    \foreach \x in {0, 1, ..., 15}
    {
        \draw (\x, 16.5) rectangle (\x + 1, 24.5);
        \node [rotate=90] at (\x + 0.5, 26.2) {\small B$_{\x, 0..7}$ };
    }
    \foreach \x in {0, 4, 8, 12}
        \draw [very thick] (\x, 16.5) rectangle (\x + 4, 24.5);

    \filldraw [fill=yellow, draw=black] (0, 28) rectangle (4, 29.5);
    \node at (2, 28.7) {T$_{0..3}$};
    \filldraw [fill=yellow, draw=black] (4, 28) rectangle (8, 29.5);
    \node at (6, 28.7) {T$_{16..19}$};
    \filldraw [fill=lime, draw=black] (8, 28) rectangle (12, 29.5);
    \node at (10, 28.7) {T$_{8..11}$};
    \filldraw [fill=lime, draw=black] (12, 28) rectangle (16, 29.5);
    \node at (14, 28.7) {T$_{24..27}$};
    \filldraw [fill=MyBlueLight, draw=black] (0, 29.5) rectangle (4, 31);
    \node at (2, 30.2) {T$_{4..7}$};
    \filldraw [fill=MyBlueLight, draw=black] (4, 29.5) rectangle (8, 31);
    \node at (6, 30.2) {T$_{20..23}$};
    \filldraw [fill=pink, draw=black] (8, 29.5) rectangle (12, 31);
    \node at (10, 30.2) {T$_{12..15}$};
    \filldraw [fill=pink, draw=black] (12, 29.5) rectangle (16, 31);
    \node at (14, 30.2) {T$_{28..31}$};

    \draw [dotted] (-8 - 0.5, 16) grid (-0.5, 0);
    \foreach \count in {0, 1, ..., 15}
    {
        \draw (-8.5, 16 - \count) rectangle (-0.5, 16 - \count - 1);
        \node at (-10.5, 16 - \count - 0.5) {\small {A$_{\count, 0..7}$} };
    }
    \foreach \y in {16, 12, 8, 4}
        \draw [very thick] (-8.5, \y) rectangle (-0.5, \y - 4);

    \filldraw [fill=yellow, draw=black] (-14, 12) rectangle (-12.5, 16);
    \node at (-13.3, 15) {T$_{0}$};
    \node at (-13.3, 14) {$..$};
    \node at (-13.3, 13) {T$_{3}$};
    \filldraw [fill=lime, draw=black] (-15.5, 12) rectangle (-14, 16);
    \node at (-14.8, 15) {T$_{8}$};
    \node at (-14.8, 14) {$..$};
    \node at (-14.8, 13) {T$_{11}$};

    \filldraw [fill=MyBlueLight, draw=black] (-14, 8) rectangle (-12.5, 12);
    \node at (-13.3, 11) {T$_{4}$};
    \node at (-13.3, 10) {$..$};
    \node at (-13.3, 9) {T$_{7}$};
    \filldraw [fill=pink, draw=black] (-15.5, 8) rectangle (-14, 12);
    \node at (-14.8, 11) {T$_{12}$};
    \node at (-14.8, 10) {$..$};
    \node at (-14.8, 9) {T$_{15}$};

    \filldraw [fill=yellow, draw=black] (-14, 4) rectangle (-12.5, 8);
    \node at (-13.3, 7) {T$_{16}$};
    \node at (-13.3, 6) {$..$};
    \node at (-13.3, 5) {T$_{19}$};
    \filldraw [fill=lime, draw=black] (-15.5, 4) rectangle (-14, 8);
    \node at (-14.8, 7) {T$_{24}$};
    \node at (-14.8, 6) {$..$};
    \node at (-14.8, 5) {T$_{27}$};

    \filldraw [fill=MyBlueLight, draw=black] (-14, 0) rectangle (-12.5, 4);
    \node at (-13.3, 3) {T$_{20}$};
    \node at (-13.3, 2) {$..$};
    \node at (-13.3, 1) {T$_{23}$};
    \filldraw [fill=pink, draw=black] (-15.5, 0) rectangle (-14, 4);
    \node at (-14.8, 3) {T$_{28}$};
    \node at (-14.8, 2) {$..$};
    \node at (-14.8, 1) {T$_{31}$};

    \newcounter{col}
    \newcommand\setrow[9]{
        \foreach \y in {#9}
        {
            \setcounter{col}{0.5}
            \foreach \n in {#1, #2, #3, #4, #5, #6, #7, #8} {
                \edef\x{2 * \value{col} + 1}
                \node[anchor=center] at (\x, \y) { \n };
                \stepcounter{col}
            }
        }
    }

    \filldraw [fill=yellow, draw=yellow] (0, 4) rectangle (8, 8);
    \filldraw [fill=yellow, draw=yellow] (0, 12) rectangle (8, 16);

    \filldraw [fill=lime, draw=lime] (8, 4) rectangle (16, 8);
    \filldraw [fill=lime, draw=lime] (8, 12) rectangle (16, 16);

    \filldraw [fill=MyBlueLight, draw=MyBlueLight] (0, 0) rectangle (8, 4);
    \filldraw [fill=MyBlueLight, draw=MyBlueLight] (0, 8) rectangle (8, 12);

    \filldraw [fill=pink, draw=pink] (8, 0) rectangle (16, 4);
    \filldraw [fill=pink, draw=pink] (8, 8) rectangle (16, 12);

    \draw[xscale=2, yscale=1] (0, 0) grid (8, 16);
    \draw[very thick, xscale=4, yscale=4] (0, 0) grid (4, 4);

\footnotesize
    \setrow {0}{2}{0}{2}{8}{10}{8}{10}{15.5}
    \setrow {1}{3}{1}{3}{9}{11}{9}{11}{14.5}
    \setrow {0}{2}{0}{2}{8}{10}{8}{10}{13.5}
    \setrow {1}{3}{1}{3}{9}{11}{9}{11}{12.5}

    \setrow {4}{6}{4}{6}{12}{14}{12}{14}{11.5}
    \setrow {5}{7}{5}{7}{13}{15}{13}{15}{10.5}
    \setrow {4}{6}{4}{6}{12}{14}{12}{14}{9.5}
    \setrow {5}{7}{5}{7}{13}{15}{13}{15}{8.5}

    \setrow {16}{18}{16}{18}{24}{26}{24}{26}{7.5}
    \setrow {17}{19}{17}{19}{25}{27}{25}{27}{6.5}
    \setrow {16}{18}{16}{18}{24}{26}{24}{26}{5.5}
    \setrow {17}{19}{17}{19}{25}{27}{25}{27}{4.5}

    \setrow {20}{22}{28}{30}{20}{22}{28}{30}{3.5}
    \setrow {21}{23}{29}{31}{21}{23}{29}{31}{2.5}
    \setrow {20}{22}{28}{30}{20}{22}{28}{30}{1.5}
    \setrow {21}{23}{29}{31}{21}{23}{29}{31}{0.5}
\small

\normalsize
\end{tikzpicture}}
\caption{\small 16$\times$16$\times$8 macro-MMA for row-major matrix A and column-major matrix B. \normalsize}
\label{fig:mma-m16n16k8-row-col}
\vspace{-0.3cm}
\end{figure}
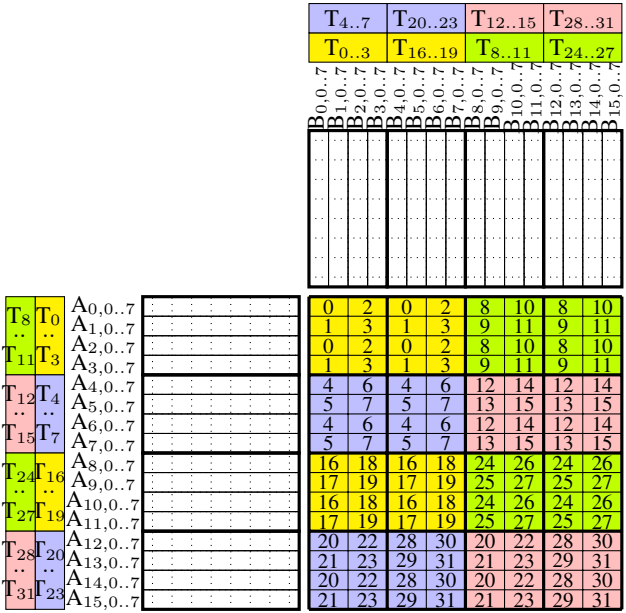

\footnotesize
\begin{figure*}[t!]
  \begin{minipage}{0.48\linewidth}
    \footnotesize DOMAIN : $S[i, j, k] : 0 \le i < M \land 0 \le j < N \land 0 \le k < K$ \\
      \hspace*{0.3cm}BAND: $S[i, j, k] \rightarrow [i, j, k]$ \\
    \vspace{-0.4cm}
    \caption{\small Initial schedule for matmul in schedule tree form.}
    \label{fig:sched1-initial}
    \vspace{0.5cm}
    DOMAIN: $S[i, j, k] : 0 \le i < M \land 0 \le j < N \land 0 \le k < K$ \\
      \hspace*{0.2cm}BAND: $S[i, j, k] \rightarrow [\lfloor{i/128}\rfloor, \lfloor{j/128}\rfloor, \lfloor{k/32}\rfloor]$ \\
      \vspace{-0.2cm}
      \mybox{\textsc{\hspace*{0.2cm}\hspace*{0.2cm}BAND: $S[i, j, k] \rightarrow [\lfloor{i/64}\rfloor - 2\lfloor{i/128}\rfloor, \lfloor{j/64}\rfloor-2\lfloor{j/128}\rfloor, 0]$ \\
      \hspace*{0.2cm}\hspace*{0.2cm}\hspace*{0.2cm}BAND: $S[i, j, k] \rightarrow [i - 64\lfloor{i/64}\rfloor, j - 64\lfloor{j/64}\rfloor, k - 32\lfloor{k/32}\rfloor]$}}\\
    \vspace{-0.15cm}
    \caption{\small Tiling with block tile size of 128$\times$128$\times$32 and warp-level tile size of 64$\times$64$\times$32.}
    \label{fig:sched1-warp-tile}
    \vspace{0.45cm}
    DOMAIN: $S[i, j, k] : 0 \le i < M \land 0 \le j < N \land 0 \le k < K$ \\
      \vspace{-0.2cm}
      \mybox{\textsc{\hspace*{0.3cm}BAND: $S[i, j, k] \rightarrow [blockIdx.y, blockIdx.x, \lfloor{k/32}\rfloor]$ \\
      \hspace*{0.3cm}\hspace*{0.3cm}BAND: $S[i, j, k] \rightarrow [warpIdx\_y, warpIdx\_x, 0]$}} \\
      \hspace*{0.3cm}\hspace*{0.3cm}\hspace*{0.3cm}BAND: $S[i, j, k] \rightarrow [i - 64\lfloor{i/64}\rfloor, j - 64\lfloor{j/64}\rfloor, k - 32\lfloor{k/32}\rfloor]$\\
    \vspace{-0.4cm}
    \caption{\small Introducing kernel parameters --
    \textit{blockIdx.y, blockIdx.x, warpIdx\_y, warpIdx\_x}.}
    \label{fig:sched1-parametric}
  \end{minipage}
  \begin{minipage}{0.02\linewidth}
    \hspace*{0.2cm}
  \end{minipage}
  \begin{minipage}{0.48\linewidth}
    \footnotesize DOMAIN: $S[i, j, k] : 0 \le i < M \land 0 \le j < N \land 0 \le k < K$ \\
      \hspace*{0.3cm}BAND: $S[i, j, k] \rightarrow [blockIdx.y, blockIdx.x, \lfloor{k/32}\rfloor]$ \\
      \hspace*{0.3cm}\hspace*{0.3cm}BAND: $S[i, j, k] \rightarrow [warpIdx\_y, warpIdx\_x, 0]$ \\
      \vspace{-0.2cm}
      \mybox{\textsc{\hspace*{0.3cm}\hspace*{0.3cm}\hspace*{0.3cm}BAND: $S[i, j, k] \rightarrow [0, 0, \lfloor{k/8}\rfloor - 4\lfloor{k/32}\rfloor]$\\
      \hspace*{0.3cm}\hspace*{0.3cm}\hspace*{0.3cm}\hspace*{0.3cm}BAND: $S[i, j, k] \rightarrow [i - 64\lfloor{i/64}\rfloor, j - 64\lfloor{j/64}\rfloor, k - 8\lfloor{k/8}\rfloor]$}}\\
    \vspace{-0.15cm}
    \caption{\small Strip-mining along the k dimension with strip-size 8.}
    \vspace{0.45cm}
    \label{fig:sched1-strip-mined}
    \mybox{\textsc{DOMAIN: $S[i, j, k] : 0 \le i < M \land 0 \le j < N \land 0 \le k < K \land 16\lfloor{i/16}\rfloor = i \land 16\lfloor{j/16}\rfloor = j \land 8\lfloor{k/8}\rfloor = k$}} \\
      \hspace*{0.3cm}BAND: $S[i, j, k] \rightarrow [blockIdx.y, blockIdx.x, \lfloor{k/32}\rfloor]$ \\
      \hspace*{0.3cm}\hspace*{0.3cm}BAND: $S[i, j, k] \rightarrow [warpIdx\_y, warpIdx\_x, 0]$ \\
      \hspace*{0.3cm}\hspace*{0.3cm}\hspace*{0.3cm}BAND: $S[i, j, k] \rightarrow [0, 0, \lfloor{k/8}\rfloor - 4\lfloor{k/32}\rfloor]$\\
      \hspace*{0.3cm}\hspace*{0.3cm}\hspace*{0.3cm}\hspace*{0.3cm}BAND: $S[i, j, k] \rightarrow [i - 64\lfloor{i/64}\rfloor, j - 64\lfloor{j/64}\rfloor, k - 8\lfloor{k/8}\rfloor]$\\
    \vspace{-0.4cm}
    \caption{\small Constraining the schedule domain so that each macro-MMA instance is mapped to a single integer point. }
    \vspace{0.1cm}
    \label{fig:sched1-constrained-domain}
  \end{minipage}
    \vspace{-0.2cm}
  \caption{\small Schedule tree transformations for matmul.}
\end{figure*}
\normalsize

Consider Figure~\ref{fig:mma-m16n16k8-row-col} which illustrates
the 16$\times$16$\times$8 macro-MMA where each warp performs a
16$\times$16$\times$8 matrix multiply-and-accumulate operation. The input
matrices $A$ and $B$ are of size 16$\times$8 and 8$\times$16 respectively.
The accumulator matrix is of size 16$\times$16. As with \textit{mma.m8n8k4},
the input matrices may be in row or column major layout, which determines how
the input matrices are distributed across the threads.
Figure~\ref{fig:mma-m16n16k8-row-col} shows the distribution for a scenario
where the input matrix A is row-major and matrix B is column-major. We refer
to the subset of data elements owned by a given thread as a \textit{macro-MMA
fragment}. Each input macro-MMA fragment consists of 8 \textit{fp16} values
-- for example, thread 0 owns the 8 elements in the first row of matrix A as well
as the 8 elements in the first column of B (note that these elements are also
owned by threads 8 and 4 respectively). The accumulator fragment
consists of 8 \textit{fp32} values. In Figure~\ref{fig:mma-m16n16k8-row-col},
the numbers inside the cells indicate the threads that own the corresponding
elements in the output matrix. For example, thread 0 owns the first two
elements in the first row as well as 3 other pairs of adjacent
elements, all of which make up its accumulator fragment. The 4 different
colours represent the 4 quad-pairs, thereby indicating the thread-to-data
mapping for all the matrices. Regardless of the layout of the input
matrices, the distribution of the accumulator matrix for the macro-MMA
remains the same.

In case of 32$\times$32$\times$8 macro-MMA an accumulator matrix of
size 32$\times$32 is computed using input matrices of size 32$\times$8 and
8$\times$32. An input macro-MMA fragment for a 32$\times$32$\times$8
macro-MMA contains 16 $fp16$ values, and 32 $fp32$ values in its accumulator
fragment. Both of these macro-MMA compositions along with wide loads and
stores can lead to performance comparable with that of hand-tuned kernels. In
the following sections, without any loss of generality, we discuss our
approach using 16$\times$16$\times$8 macro-MMA as the reference. A similar
approach holds good for 32$\times$32$\times$8 macro-MMA as well.

\subsection{Tiling for Blocks and Warps}

The outer-parallel schedule obtained using ISL is shown in
Figure~\ref{fig:sched1-initial}, in the schedule tree form. It corresponds to
a 3-d loop-nest and consists of a single permutable band with 3 schedule
dimensions. Only the innermost dimension is sequential.

Suppose we distribute the loop-nest across a 2-d grid of size
$b_1\times{b_2}$ with each thread block having a
2-d $w_1\times w_2$ arrangement of warps. Furthermore, suppose the
sequential loop is to be strip-mined with a strip size of $b_s$. This can be
achieved by successively tiling the loop-nest, first with tile sizes
${b_1}\times{b_2}\times{b_s}$ and then with tile sizes
${w_1}\times{w_2}\times{b_s}$. We refer to the former as the \textit{block
tile} and the latter as the \textit{warp tile}. 
Clearly, the block tile sizes along every dimension must exceed or
equal the warp tile sizes along the corresponding dimension. Also,
they must be integer multiples of the macro-MMA sizes chosen --
16$\times$16$\times$8 or 32$\times$32$\times$8. The
problem sizes are assumed to be multiples of the block tile sizes. For other
problem sizes, we can either pad the input matrices with zeroes to make them
so or generate specialized code for the partial tiles and then combine the
results of the full-tile computed using macro-MMA and the results of the
partial-tiles. 

Figure~\ref{fig:sched1-warp-tile} shows the block and warp-level tiling
transformation on the schedule tree using tile sizes 128$\times$128$\times$32
and 64$\times$64$\times$32 respectively. Note that in all three band nodes of
the resulting schedule tree, only the innermost dimension is sequential.
Clearly, the parallel dimensions of the outermost band correspond to the
block indices, \textit{blockIdx.y} and \textit{blockIdx.x}. Similarly, the
middle band essentially iterates over the warp tiles. So, its parallel
dimensions correspond to warp indices, \textit{warpIdx\_y} and
\textit{warpIdx\_x}, which can be derived from other kernel
parameters such as thread indices. Figure~\ref{fig:sched1-parametric} shows
the schedule tree after these schedule parameters have been introduced.

\subsection{Strip Mining the Warp-Level MMA}

The innermost band in Figure~\ref{fig:sched1-parametric}, which performs a
warp-level MMA operation, processes a warp tile of size
64$\times$64$\times$32. As shown in Figure~\ref{fig:sched1-strip-mined}, the
sequential dimension in the innermost band can be strip-mined further using a
strip size of 8, which corresponds to the macro-MMA size along the $k$
dimension. Such a restructuring essentially expresses the warp-level MMA as
an outer product, which exposes more instruction-level parallelism than an
inner product formulation. This helps cover the latency from instructions and
the memory load at a low occupancy.


\subsection{Schedule Domain Contraction}
\label{sec:sched1-schedule-dom-transformation}

At this stage, the innermost band in Figure~\ref{fig:sched1-strip-mined}
specifies a 3-d loop nest around the statement $S$, each instance of which
performs a scalar matrix-multiply and accumulate operation. To target tensor
cores, each statement instance must instead perform a macro-MMA operation (of
shape 16$\times$16$\times$8 or 32$\times$32$\times$8). This is done by
constraining the schedule domain as shown in
Figure~\ref{fig:sched1-constrained-domain}, where an entire subset of integer
points corresponding to a 16$\times$16$\times$8 macro-MMA is mapped to a
single integer point. The root operation in the expression tree associated with the
statement $S$ is then altered to one that performs a 16$\times$16$\times$8
macro-MMA. This means that the innermost band of the schedule tree would then
specify the schedule for a loop iterating over 16$\times$16$\times$8
macro-MMA instances instead of scalar multiply-accumulate operations. In
effect, this distributes the warp-level iteration space across all the
threads of the warp. Note that the updated operation type can be of four
different types -- one for each of the macro-MMA layout specializations.

At this stage, we assume that the macro-MMA fragments owned by each thread
are somehow available for the macro-MMA operation. In later sections, we
discuss how these fragments can be created and loaded with the required data.

\footnotesize
\begin{figure*}[t!]
  \begin{minipage}{0.48\linewidth}
    \footnotesize DOMAIN: $S[i, j, k] : 0 \le i < M \land 0 \le j < N \land 0 \le k < K$ \\
      \hspace*{0.1cm}BAND: $S[i, j, k] \rightarrow [\lfloor{i/128}\rfloor, \lfloor{j/128}\rfloor, \lfloor{k/64}\rfloor]$ \\
      \vspace{-0.2cm}
      \mybox{\textsc{\hspace*{0.1cm}\hspace*{0.1cm}BAND: $S[i, j, k] \rightarrow [\lfloor{i/64}\rfloor - 2\lfloor{i/128}\rfloor, \lfloor{j/64}\rfloor-2\lfloor{j/128}\rfloor,$ $\lfloor{k/32}\rfloor-2\lfloor{k/64}\rfloor]$ \\
      \hspace*{0.1cm}\hspace*{0.1cm}\hspace*{0.2cm}BAND: $S[i, j, k] \rightarrow [i - 64\lfloor{i/64}\rfloor, j - 64\lfloor{j/64}\rfloor,k - 32\lfloor{k/32}\rfloor]$}}\\
    \vspace{-0.05cm}
    \caption{\small Tiling initial schedule in
    Figure~\ref{fig:sched1-initial} with block tile size of
    128$\times$128$\times$64 instead of 128$\times$128$\times$32 for 2-way
    split-K. Warp tile size remains the same --
    64$\times$64$\times$32.}
    \label{fig:sched2-warp-tile}
  \end{minipage}
  \begin{minipage}{0.02\linewidth}
    \hspace*{0.2cm}
  \end{minipage}
  \begin{minipage}{0.48\linewidth}
    \footnotesize DOMAIN: $S[i, j, k] : 0 \le i < M \land 0 \le j < N \land 0 \le k < K$ \\
      \vspace{-0.2cm}
      \mybox{\textsc{\hspace*{0.3cm}BAND: $S[i, j, k] \rightarrow [blockIdx.y, blockIdx.x, \lfloor{k/64}\rfloor]$ \\
      \hspace*{0.3cm}\hspace*{0.3cm}BAND: $S[i, j, k] \rightarrow [warpIdx\_y, warpIdx\_x, warpIdx\_z]$}} \\
      \hspace*{0.3cm}\hspace*{0.3cm}\hspace*{0.3cm}BAND: $S[i, j, k] \rightarrow [i - 64\lfloor{i/64}\rfloor, j - 64\lfloor{j/64}\rfloor, k - 32\lfloor{k/32}\rfloor]$\\
    \vspace{-0.3cm}
    \caption{\small On introducing kernel parameters --
    \textit{blockIdx.y, blockIdx.x, warpIdx\_y, warpIdx\_x, warpIdx\_z}.}
    \label{fig:sched2-parametric}
  \end{minipage}
  \caption{\small Schedule tree transformations for matmul with 2-way intra-thread-block split-K.}
\end{figure*}
\normalsize

\subsection{Split-K}

The compute decomposition described so far parallelizes the matmul
computation only along the parallel dimensions, i.e., the $i$ and $j$ loops.
While the $k$ dimension is sequential, it is reasonable to parallelize along
the $k$ dimension as well when targeting tensor cores. We refer to such a
parallelization as \textit{split-K}. A 2-way \textit{intra-thread-block split-K}
splits the computation into 2 parts and assigns them to concurrent warps
along the $z$ axis of the GPU compute hierarchy. The partial results computed
by corresponding warps along the $z$ axis then need to be summed up to
obtain the final result for the matmul computation. More generally, it is
possible to extend this arrangement to a \textit{p}-way intra-thread-block split-K,
where $p$ is a power of 2.

\subsubsection{Tiling for Intra-Thread-Block Split-K}
For a $p$-way split-K, the block tiles need to be bigger by a factor $p$
along the $k$ dimension, while the warp tiles are of the same size as before.
So, the compute decomposition for a $p$-way split-K involves tiling the
loop-nest, first with tile sizes ${b_1}\times{b_2}\times{p*b_s}$ and then with
${w_1}\times{w_2}\times{b_s}$. This creates $p$ rows of warp tiles along the
$z$ axis of the GPU compute hierarchy. Figure ~\ref{fig:sched2-warp-tile}
illustrates this approach for a 2-way intra-thread-block split-K. Clearly, the
resulting schedule in the middle band is for a loop-nest that iterates over
warp tiles along the $y$, $x$ and $z$ dimensions respectively. So, on
introducing these derived kernel parameters, along with the block indices,
the schedule tree would be as shown in Figure~\ref{fig:sched2-parametric}.

The above tilings are followed up with strip-mining
and schedule domain contraction as shown earlier in
Figures~\ref{fig:sched1-strip-mined} and~\ref{fig:sched1-constrained-domain}
to complete the compute decomposition for a $p$-way split-K.

\section{Data Movement across Memory Hierarchy}
\label{sec:core2}

In order to obtain good performance on a GPU, it is important to optimize the
movement of data across global memory, shared memory and registers. In this
section, we explain how copy statements that move the data across this memory
hierarchy are introduced by inserting their associated copy schedule nodes to
the transformed schedule tree obtained after compute decomposition.

The read maps, $I\rightarrow M_A$ and $I\rightarrow M_B$ capture the read
access relation from the iteration domain of the matmul statement $S$ to the
dataspaces $M_A$ and $M_B$, which are only read. The schedule tree
transformation described in the previous section effectively maps the
iteration space $I$ to the schedule space $I'$ with the schedule vector
$(b_y, b_x, c_0, w_y, w_x, c_1, c_2, c_3, c_4, c_5, c_6, c_7)$. The dimensions
$b_y, b_x, c_0$ correspond to the schedule dimensions of the outermost band
in Figure~\ref{fig:sched1-constrained-domain}; $w_y, w_x, c_1$ correspond to
the schedule dimensions of the band immediately below it and so on.

\subsubsection{Global, Shared and Register Data Tiles}
Using the transformation $I\rightarrow{I}'$, we can derive the mappings
$I'\rightarrow M_A$, $I'\rightarrow M_B$ and likewise, the write mapping $I'
\rightarrow M_C$, which describe the read and the write access relations from
the schedule space $I'$ to the input and output dataspaces respectively. The
basic idea is to infer tiles in the global memory that are accessed at
block-level and warp-level. Global data tiles that are reused can then be
promoted to shared memory data tiles and register tiles, i.e., copied to
shared memory or register files and then reused. For each memory 
promotion, a new statement is introduced to perform this data copy.
Furthermore, the access maps can be analyzed to determine the access depth as
well as the dimensions of these data tiles. The iteration domain of the copy
statement can be derived from the data space of the corresponding data tile.
At the inferred access depth, a new schedule node defining a schedule for the
copy statement is inserted into the schedule tree.
As explained later, in some cases, the copy operation may
not be a simple data assignment and may involve a call to helper functions.

\subsection{Global to Shared Memory}
\label{subsec:global-to-shared}

Block-level read maps $I'_{block}\rightarrow M_A$ and $I'_{block}\rightarrow
M_B$ can be derived from the read access relations $I'\rightarrow M_A$ and
$I'\rightarrow M_B$ respectively, by projecting out all the dimensions other
than those in the outer schedule band, namely, $(b_y, b_x, c_0)$. The ranges
of these maps determine the data tiles that are accessed by the compute tile
$(b_y, b_x, c_0)$. Since this data tile is reused by threads within the block
with block indices $(b_y, b_x)$, it can be promoted to shared memory.



Furthermore, we separate out the global-to-shared copy by splitting it into a
global-to-register copy followed by a register-to-shared copy loop. The
global-to-register copy loop is set up in such a way that each thread
performs 128-bit accesses (8 contiguous fp16 elements) to global memory.
Furthermore, this loop to copy data from global memory to registers using
vectorized loads is cyclically distributed across all the threads in a thread block.
Such a distribution also ensures that global accesses are coalesced.


\subsubsection{Swizzled Shared Memory}
In order to ensure that there are no bank conflicts when reading and writing
data to and from the shared memory, the data elements in the shared memory
tile need to be permuted or \textit{swizzled}. So, unlike the copy from
global memory to registers, which is a straightforward copy of an 8-wide
vector of \textit{fp16} data values, the copy from registers to shared memory
involves the application of a swizzle function. In essence, the 8-wide vector
data is stored to a shared memory buffer where the data ordering is different
from that in the corresponding global data tile.

The swizzle function, used to compute the offset in the swizzled shared
memory allocation, provides a one-to-one mapping from the unswizzled data
tile to the swizzled tile in shared memory. Essentially, it maps the offsets
of contiguous blocks of 8 \textit{fp16} data values in the unswizzled data tile
to an offset in the shared memory allocation for that data tile. The exact
implementation of the swizzle function depends on the number of elements per
row and column in the swizzled shared array.
We also use different helper functions to copy the
data to shared memory for each of the 4 combinations -- $A$ or $B$ in row or
column major layout.

\subsection{Shared Memory to Register Fragments}

The access maps $I'_{warp}\rightarrow M_A$, $I'_{warp}\rightarrow M_B$,
$I'_{warp}\rightarrow M_C$ are obtained by projecting out the dimensions
$(c_5, c_6, c_7)$ from $I'\rightarrow M_A$, $I'\rightarrow M_B$,
$I'\rightarrow M_C$ respectively. The ranges of these warp-level access maps
determine the input and output data tiles for the warp-level macro-MMA. For
the schedule tree in Figure~\ref{fig:sched1-constrained-domain}, these data
tiles would be of size 64$\times$8, 8$\times$64 and 64$\times$64 respectively.

\subsubsection{Macro-MMA Register Fragments}

Each thread need not own the entire data tile. Recall the schedule domain
contraction in Figure~\ref{fig:sched1-constrained-domain} which abstracted
away the inner point loops. Similarly, these tiles can be contracted
to obtain an array of register fragments. For a 16$\times$16$\times$8
macro-MMA, the sizes of the array of macro-MMA fragments can be found as
follows.

\begin{itemize}
    \item The 2-d input data tile of $M_A$ is contracted by factors 16$\times$8
    if it is row major, and by factors 8$\times$16 if it is column major.
    For example, a given tile of size 64$\times$8 can
    contracted down to a 4$\times$1 array of register fragments.
    
    \item The input data tile of $M_B$ is contracted by factors 8$\times$16
    if it is row major and by sizes 16$\times$8 if it is column major.
    
    \item The output data tile for $M_C$ is contracted by factors 16$\times$16
    to obtain the array of accumulator fragments.
\end{itemize}

Each 16$\times$16$\times$8 macro-MMA fragment contains 8 \textit{fp16}
elements.


\subsubsection{Loading Register Fragments}
In order to load the register fragments, a polyhedral schedule can be created
for moving the data from shared memory to register fragments. Each register
fragment load involves loading 8 \textit{fp16} values from shared memory. The
swizzled storage ensures that there are no bank conflicts due to the read
accesses to it. The same swizzling function that is used for storing to shared
memory is used to obtain the swizzled offset from which to fetch the data
vector in the swizzled storage. Again, the data copy is done using
vectorized accesses. The macro-MMA thread-to-data mappings for inputs $A$ and
$B$ are different. Furthermore, it is different for different layouts -- row
or column major. Also, different load fragment helper functions are needed
for each of the 4 possibilities -- $A$ or $B$ in row or column major layout.


\begin{figure*}
\lstset{style=CUDA}
\begin{lstlisting}[mathescape=true, caption={\small Skeleton code for the matmul kernel generated with block tile size 128$\times$128$\times$32 and warp tile size 64$\times$64$\times$32. \normalsize},label={lst:matmul-cuda}]
extern "C" __global__ void __launch_bounds__(256) kern0(int M, int N, int K, const half * __restrict__ M_0, int ldM_0, const half * __restrict__ M_1, int ldM_1, half * __restrict__ M_3, int ldM_3) {
  ...
  for (int c2 = -1; c2 < K / 64; c2 += 1) // 2-way split-K
    if (K >= 64 * c2 + 128) // prefetch data from global memory
      #pragma unroll for (int c5 = 0; c5 <= 3; c5 += 1) {
        (half8&)(private_M_1[(8 * c5)]) = ((half8&)(M_0[(128 * blockIdx.y + 32 * c5 + linearId / 8) * K + (8 * (linearId % 8) + 64 * c2 + 64) * 1])); // copy data from global memory to registers
        (half8&)(private_M_3[(8 * c5)]) = ((half8&)(M_1[(128 * blockIdx.x + 32 * c5 + linearId / 8) * K + (8 * (linearId % 8) + 64 * c2 + 64) * 1]));} // copy data from global memory to registers
    if (c2 >= 0) // overlap computation with data movement from global memory to registers
      #pragma unroll for (int c8 = 0; c8 <= 3; c8 += 1){ // strip-mine the warp-level macro-MMA
        #pragma unroll for (int c11 = 0; c11 <= 3; c11 += 1){
          hmma_load_b_col_swizzled(&hmma_M_5[(c11)][(0)][0][0], &shared_mma_M_2[0], (64 * warpIdx_x + 16 * c11), (32 * warpIdx_z + 8 * c8), 64); // copy data from swizzled shared buffers to register fragments
          hmma_load_a_row_swizzled(&hmma_M_4[(c11)][(0)][0][0], &shared_mma_M_0[0], (64 * warpIdx_y + 16 * c11), (32 * warpIdx_z + 8 * c8), 64);} // copy data from swizzled shared buffers to register fragments
        // iterate over the macro-MMAs
        #pragma unroll for (int c9 = 0; c9 <= 63; c9 += 16){
          #pragma unroll for (int c10 = 0; c10 <= 63; c10 += 16){ // perform macro-MMA (A is row & B is col-major)
            hmma_row_col(&hmma_M_6[(c9 / 16)][(c10 / 16)][0][0], &hmma_M_4[(c9 / 16)][(0)][0][0], &hmma_M_5[(c10 / 16)][(0)][0][0], &hmma_M_6[(c9 / 16)][(c10 / 16)][0][0]);}}}
    if (K >= 64 * c2 + 128) // prefetch data from shared memory
      __syncthreads();
      #pragma unroll for (int c5 = 0; c5 <= 3; c5 += 1) { // copy data from registers to swizzled shared buffers
        hmma_store_shared_b_col_swizzled(&shared_mma_M_2[0], ((half8&)(private_M_3[(8 * c5)])), (8 * c5));
        hmma_store_shared_a_row_swizzled(&shared_mma_M_0[0], ((half8&)(private_M_1[(8 * c5)])), (8 * c5));}
      __syncthreads();
  if (K >= 64) // iterate over accumulator fragments and store results to global mem, rearrange through shared mem
    #pragma unroll for (int c4 = 0; c4 <= 3; c4 += 1)
      #pragma unroll for (int c5 = 0; c5 <= 3; c5 += 1)
        hmma_store_global_after_reordering(&M_3[(64 * warpIdx_y + 128 * blockIdx.y + 16 * c4) * N + (64 * warpIdx_x + 128 * blockIdx.x + 16 * c5) * 1], &hmma_M_6[(c4)][(c5)][0][0], ldM_3, &sharedBuffer[0]);
}
\end{lstlisting}
\vspace{-0.3cm}
\end{figure*}

\subsection {Macro-MMA on Register Fragments}
As described in the previous subsection, the register fragments are inferred
from the access maps $I'_{warp}\rightarrow M_A$, $I'_{warp}\rightarrow M_B$,
$I'_{warp}\rightarrow M_C$. Consequently, a mapping from the global dataspace
to the data space of register fragments can be inferred. Using these
mappings, all accesses to global memory in the statement $S$ can be altered
to access register fragments instead.

\subsection{Store to Global Memory}
\label{sec:store-global}

The result of a macro-MMA computation goes into an accumulator fragment,
which resides in registers. So, a copy of the data from accumulator fragments
to global memory is required.

\subsubsection{Reordering through Shared Memory}

The 8 \textit{fp32} data values making up an accumulator fragment do not map
to a contiguous block of 8 elements (please refer
Figure~\ref{fig:mma-m16n16k8-row-col}). Furthermore, it is necessary to
convert these data values from the \textit{fp32} accumulator type to the
\textit{fp16} output type. So, this precludes a straightforward 8-wide
vectorized store to global memory.
In order to achieve that, the overall strategy is to move the data from the
accumulator fragments to global memory, via shared memory. Essentially,
shared memory is used as a temporary buffer to exchange and rearrange the
accumulator data in registers so that each thread can finally move a
contiguous block of 8 \textit{fp16} values to global memory. For example, as
shown in Figure~\ref{fig:mma-m16n16k8-row-col}, thread 0 owns the first two
elements $C_{0,0..1}$ in the first row of the accumulator matrix, as well as
the elements $C_{0,4..5}$. Likewise, thread 2 owns $C_{0,2..3}$ and
$C_{0,6..7}$. Thread 0 packs the 4 elements that it owns into a 4-wide vector
of $fp16$ values (after converting them from \textit{fp32} to \textit{fp16}),
and copies them to a contiguous block in shared memory. Now, if thread 2 does
the same and copies its packed data to the adjacent contiguous block of
shared memory, thread 0 can then copy back the entire contiguous block of 8
\textit{fp16} values to registers and rearrange them to obtain the correct
data order. A similar procedure is employed for all the threads to make sure
that the threads own data elements that form contiguous blocks in the global
array. Again, the accesses to shared memory are modeled to avoid bank
conflicts and to exploit wide loads and stores.

\subsubsection{Split-K Reduction}
In case of a split-K schedule, the partial results computed by corresponding
warps along the $z$ axis must be summed up. This is done by adding the
partial results once they are stored to shared memory for the reordering 
described above. The reduction is performed only by the threads in warps with
$warp\_z$ equal to 0. Furthermore, this means that the rest of the threads do
not perform any store to global memory.


Reordering involves carefully chosen non-affine access patterns. So, all
these details, including data conversion and split-K reduction, are hidden
behind a helper function that copies data from an accumulator fragment to
global memory through the above steps.

\subsection{Prefetching and Hiding Latency}



As explained earlier, the input data for the macro-MMA computation is
obtained by copying the data from global memory to shared memory and then
from the latter to the register fragments. A prefetching schedule can be
obtained by shifting the schedule bands associated with the corresponding
copy schedule nodes accordingly. The schedule is further modified so that the
global memory access latency is hidden by overlapping it with the macro-MMA
computation.

\footnotesize
\begin{figure*}[t!]
  \begin{minipage}{0.48\linewidth}
    \lstset{linewidth=0.98\linewidth}
    \begin{lstlisting}[mathescape=true, caption={Matmul + Bias + ReLU},label={lst:matmul-bias-relu-scheduled}]
for(i = 0; i $<$ M; ++i)
 for(j = 0; j $<$ N; ++j) {
  for(k = 0; k $<$ K; ++k)
   C[i, j] = mul_acc(C[i, j], A[i, k], B[k, j]);//S1
  E[i, j] = relu_add(C[i, j], bias[i, j]);//S2
 }
    \end{lstlisting}
     \vspace{0.5cm}
    \footnotesize DOMAIN : $S_1[i, j, k] : 0 \le i < M \land 0 \le j < N \land 0 \le k < K;$ \\
    \hspace*{1.5cm}$S_2[i, j] : 0 \le i < M \land 0 \le j < N$ \\
      \hspace*{0.1cm}BAND: $S_1[i, j, k] \rightarrow [i, j, k]; S_2[i, j] \rightarrow [i, j, K];$ \\
      \hspace*{0.1cm}\hspace*{0.1cm}SEQUENCE\\
      \hspace*{0.1cm}\hspace*{0.1cm}\hspace*{0.2cm}FILTER: ${S_1[i, j, k]}$ \\
      \hspace*{0.1cm}\hspace*{0.1cm}\hspace*{0.2cm}FILTER: ${S_2[i, j]}$ \\
     \vspace{-0.3cm}
    \caption{\small Initial schedule tree for matmul+bias+ReLU \normalsize}
    \label{fig:sched3-initial}
     \vspace{0.5cm}
     DOMAIN : $S_1[i, j, k] : 0 \le i < M \land 0 \le j < N \land 0 \le k < K;$ \\
     \hspace*{1.5cm}$S_2[i, j] : 0 \le i < M \land 0 \le j < N$ \\
     \vspace{-0.2cm}
       \mybox{\textsc{\hspace*{0.1cm}BAND: $S_1[i, j, k] \rightarrow [\lfloor{i/128}\rfloor, \lfloor{j/128}\rfloor, \lfloor{k/32}\rfloor];$ \\
       \hspace*{1.5cm}$S_2[i, j] \rightarrow [\lfloor{i/128}\rfloor, \lfloor{j/128}\rfloor, \lfloor{K/32}\rfloor] $ \\
       \hspace*{0.1cm}\hspace*{0.1cm}BAND: $S_1[i, j, k] \rightarrow [i - 128\lfloor{i/128}\rfloor, j-128\lfloor{j/128}\rfloor, k-32\lfloor{k/32}\rfloor];$ $S_2[i, j] \rightarrow [i - 128\lfloor{i/128}\rfloor, j-128\lfloor{j/128}\rfloor, K-32\lfloor{K/32}\rfloor] $}} \\
       \vspace{0.1cm}
       \hspace*{0.4cm}SEQUENCE\\
       \hspace*{0.5cm}FILTER: ${S_1[i, j, k]}$ \\
       \hspace*{0.5cm}FILTER: ${S_2[i, j]}$ \\
     \vspace{-0.3cm}
     \caption{\small Tiling with block tile size of 128$\times$128$\times$32.\normalsize}
     \label{fig:sched3-block-tile}
  \end{minipage}
   \begin{minipage}{0.02\linewidth}
    \hspace*{0.02cm}
   \end{minipage}
   \begin{minipage}{0.48\linewidth}
     \footnotesize DOMAIN : $S_1[i, j, k] : 0 \le i < M \land 0 \le j < N \land 0 \le k < K;$ \\
     \hspace*{1.5cm}$S_2[i, j] : 0 \le i < M \land 0 \le j < N$ \\
     \vspace{-0.2cm}
       \mybox{\textsc{\hspace*{0.1cm}BAND: $S_1[i, j, k] \rightarrow [\lfloor{i/128}\rfloor, \lfloor{j/128}\rfloor];$ $S_2[i, j] \rightarrow [\lfloor{i/128}\rfloor, \lfloor{j/128}\rfloor] $ \\
       \hspace*{0.1cm}\hspace*{0.1cm}SEQUENCE\\
       \hspace*{0.3cm}FILTER: ${S_1[i, j, k]}$ \\
       \hspace*{0.4cm}BAND: $S_1[i, j, k] \rightarrow [\lfloor{k/32}\rfloor]$ \\
       \hspace*{0.5cm}BAND: $S_1[i, j, k] \rightarrow [i - 128\lfloor{i/128}\rfloor, j-128\lfloor{j/128}\rfloor, k-32\lfloor{k/32}\rfloor]$ \\
       \hspace*{0.3cm}FILTER: $S_2[i, j]$ \\
       \hspace*{0.4cm}BAND: $S_2[i, j] \rightarrow [\lfloor{K/32}\rfloor]$ \\
       \hspace*{0.5cm}BAND:$S_2[i, j] \rightarrow [i - 128\lfloor{i/128}\rfloor, j-128\lfloor{j/128}\rfloor, K-32\lfloor{K/32}\rfloor]$}} \\
     \vspace{-0.2cm}
     \caption{\small Sequence hoisting for inter-statement dependence.\normalsize}
     \label{fig:sched3-separation}
     \vspace{0.2cm}
     DOMAIN : $S_1[i, j, k] : 0 \le i < M \land 0 \le j < N \land 0 \le k < K \land 16\lfloor{i/16}\rfloor = i \land 16\lfloor{j/16}\rfloor = j \land 8\lfloor{k/8}\rfloor = k; S_2[i, j] : 0 \le i < M \land 0 \le j < N \land 16\lfloor{i/16}\rfloor = i \land 16\lfloor{j/16}\rfloor = j$ \\
     \vspace{-0.2cm}
       \mybox{\textsc{\hspace*{0.1cm}BAND: $S_1[i, j, k] \rightarrow [blockIdx.y, blockIdx.x];$ $S_2[i, j] \rightarrow [blockIdx.y, blockIdx.x]$}}  \\
       \hspace*{0.3cm}SEQUENCE\\
       \hspace*{0.4cm}FILTER: $S_1[i, j, k]$ \\
       \hspace*{0.5cm}BAND: $S_1[i, j, k] \rightarrow [\lfloor{k/32}\rfloor]$ \\
       \vspace{-0.2cm}
       \mybox{\textsc{\hspace*{0.6cm}BAND: $S_1[i, j, k] \rightarrow [warpIdx\_y, warpIdx\_x, 0];$ \\
       \hspace*{0.7cm}BAND: $S_1[i, j, k] \rightarrow [0, 0, \lfloor{k/8}\rfloor - 4\lfloor{k/32}\rfloor];$\\
       \hspace*{0.8cm}BAND: $S_1[i, j, k] \rightarrow [i - 64\lfloor{i/64}\rfloor, j - 64\lfloor{j/64}\rfloor, k - 8\lfloor{k/8}\rfloor];$}}\\
       \hspace*{0.4cm}FILTER: ${S_2[i, j]}$ \\
       \hspace*{0.5cm}BAND: $S_2[i, j] \rightarrow [\lfloor{K/32}\rfloor]$ \\
       \vspace{-0.2cm}
       \mybox{\textsc{\hspace*{0.6cm}BAND: $S_2[i, j] \rightarrow [warpIdx\_y, warpIdx\_x, 0];$ \\
       \hspace*{0.7cm}BAND: $S_2[i, j] \rightarrow [0, 0, \lfloor{K/8}\rfloor - 4\lfloor{K/32}\rfloor];$\\
       \hspace*{0.8cm}BAND: $S_2[i, j] \rightarrow [i - 64\lfloor{i/64}\rfloor, j - 64\lfloor{j/64}\rfloor, K - 8\lfloor{K/8}\rfloor];$}}\\
   \vspace*{-0.2cm}
     \caption{\small Schedule after tiling innermost bands in both the filter
     nodes with warp-level tile size of 64$\times$64$\times$32 and then
     strip-mining the sequential dimension with strip-size of 8.\normalsize}
     \label{fig:sched3-warp-tile-strip-mined}
   \end{minipage}
  \caption{\small Schedule tree transformation for Matmul + Bias + ReLU}
\end{figure*}
\normalsize

\section{Code Generation}
\label{sec:code-gen}

The AST generation facility provided by ISL is used to emit CUDA code from
the schedule tree that results from the compute and data decomposition
discussed in the previous sections. Kernel launch parameters, i.e., the grid
and block sizes are inferred through the constraints on the corresponding
schedule dimensions in the schedule tree. Listing~\ref{lst:matmul-cuda}
provides a skeleton of the kernel generated using our approach for matmul
with block tile sizes 128$\times$128$\times$32 and warp tile sizes
64$\times$64$\times$32 with a 2-way intra-thread-block split-K, where the input
matrices \texttt{M\textunderscore{0}} and \texttt{M\textunderscore{1}} are in
row-major and column major layout respectively. As can be seen the core
matmul computation is performed on line 16 through the call to the helper
function \small \texttt{hmma$\_$row$\_$col}\normalsize, which we implement
using the \textit{mma.sync.m8n8k4} instructions to realize either a
16$\times$16$\times$8 or a 32$\times$32$\times$8 macro-MMA. Similarly, the
data copy operations are implemented as calls to helper functions.
\section{Kernel Fusion}
\label{sec:core3}

We now discuss how fused kernels can be generated for some
computation idioms with both matmul and pointwise operations.

\subsection{Pointwise Operations in Epilogue}
Consider a computation sequence with pointwise operations fed by a matmul
e.g. matmul followed by a bias add and ReLU activation, with only the final
result being live out. Without loss of generality, all the pointwise
operations can be fused together and represented by a single compound operation
as in Listing~\ref{lst:matmul-bias-relu}. The ISL schedule for this is as
shown in Figure~\ref{fig:sched3-initial}, which corresponds to the loop-nest
shown in Listing~\ref{lst:matmul-bias-relu-scheduled}. The sequence node
specifies the relative ordering of the statements that appear in its filter
child nodes. Only the innermost dimensions of band nodes are sequential.
Note that unlike in Listing~\ref{lst:matmul-bias-relu}, the outer-parallel
loops are fused in Listing~\ref{lst:matmul-bias-relu-scheduled}.

As in the case of block-level tiling for just matmul, the band node can be tiled
using tile sizes $b_1\times{b_2}\times{b_s}$.
Figure~\ref{fig:sched3-block-tile} shows the result of
such a block-tiling. However, given the sequence ordering between statements
$S_1$ and $S_2$, it is necessary to hoist the sequence node further up the
schedule tree so that all sequential loop dimensions appear only in
descendant nodes of the hoisted sequence node. This may require splitting the
band members so that the outer parallel loop dimensions fall in a separate
band from that of the inner sequential dimensions.
Figure~\ref{fig:sched3-separation} illustrates this restructuring. Note that the parent band node of the sequence node only
contains parallel loop dimensions. The sequential loop dimensions for the two
statements are now moved to separate band nodes immediately under the filter
nodes. Such a restructuring ensures that the matmul reduction loop is
scheduled before its result is consumed by the pointwise operations.

Clearly, the schedule dimensions constituting the outermost band in
Figure~\ref{fig:sched3-separation} correspond to kernel parameters
\textit{blockIdx.y} and \textit{blockIdx.x}, as shown in
Figure~\ref{fig:sched3-warp-tile-strip-mined}. Furthermore, the innermost
band under both the filter nodes in Figure~\ref{fig:sched3-separation}
iterate over the warp tiles. Consequently, as shown in
Figure~\ref{fig:sched3-warp-tile-strip-mined}, we can perform warp-tiling and
strip-mining on both of these filter nodes similar to the transformations
illustrated in Figure~\ref{fig:sched1-warp-tile} and
~\ref{fig:sched1-strip-mined}. Schedule domain contraction, as described in
Section~\ref{sec:sched1-schedule-dom-transformation} is then applied for
statement $S_1$ so that it performs a 16$\times$16$\times$8 macro-MMA
operation. Similarly, for ease of code generation, the schedule domain of the
statement $S_2$ is also constrained and its operation type updated to one
that performs a pointwise operation on an entire accumulator fragment
computed by the macro-MMA operation, effectively distributing the
pointwise operation across threads.

\paragraph{Avoiding Intermediate Writes To Global Memory} The schedule space
for $S_1$ is similar to that obtained earlier for matmul alone (see
Figure~\ref{fig:sched1-constrained-domain}). So, similar memory
promotions to those described in Section~\ref{sec:core2} hold good for this
part of the schedule tree. However, since the matmul result is not live-out,
the data in an accumulator fragment need not be stored to global memory.
Nevertheless, this result is consumed by the pointwise operations scheduled
in the other filter node. So, the accumulator data needs to be converted to
\textit{fp16} type and then reordered through shared memory, but finally
copied back to a register fragment. The reordering ensures that each thread
holds a contiguous block of 8 fp16 values in every register fragment.

\paragraph{Register Fragments for Pointwise Operation} The register fragments
obtained from the accumulator data serve as one of the inputs to the
downstream pointwise operation. For the statement $S_2$, every other data
tile accessed by its warp-level compute tile is promoted to registers e.g.
those from access to the \texttt{bias} array. These data tiles are similarly
distributed across the threads by contracting them by a factor 16$\times$16
to obtain an array of register fragments, each containing 8 \textit{fp16}
elements. Note that since these fragments have the same data distribution as
the register fragments obtained from the accumulator data, they can be loaded
from global memory using 128-bit accesses. A copy schedule node is inserted
into the schedule tree for the register fragment load operations.

\paragraph{Storing Live-Out to Global Memory} With all the data tiles
accessed by a warp-level compute tile of the pointwise operation distributed
across threads into register fragments, the result of the pointwise operation
is also in a register fragment. Since it is a contiguous block of 8
\textit{fp16} values, a final copy schedule node is inserted after the
pointwise operation to perform a 128-bit store to global memory for each
register fragment.

\subsection{Matmuls with Pointwise Epilogue}
Consider the scenario where there are two matmuls of the same shape whose
results are consumed by a downstream pointwise operation. The ISL schedule
obtained for such an example is shown in Figure~\ref{sched:mm-mm-add}. The
corresponding loop-nest is shown in Listing~\ref{lst:mm-mm-add}.

\footnotesize
\begin{figure}[h!]
    \raggedright
    \footnotesize DOMAIN : $S_1[i, j, k] : 0 \le i < M \land 0 \le j < N \land 0 \le k < K;$ \\
    \hspace*{1.5cm}$S_2[i, j, k] : 0 \le i < M \land 0 \le j < N \land 0 \le k < K;$ \\
    \hspace*{1.5cm}$S_3[i, j] : 0 \le i < M \land 0 \le j < N$ \\
      \hspace*{0.2cm}BAND: $S_1[i, j, k] \rightarrow [i, j, k]; S_2[i, j, k] \rightarrow [i, j, k];$ $S_3[i, j] \rightarrow [i, j, K];$ \\
      \hspace*{0.2cm}\hspace*{0.1cm}SEQUENCE\\
      \hspace*{0.2cm}\hspace*{0.1cm}\hspace*{0.2cm}FILTER: ${S_1[i, j, k]}$ \\
      \hspace*{0.2cm}\hspace*{0.1cm}\hspace*{0.2cm}FILTER: ${S_2[i, j, k]}$ \\
      \hspace*{0.2cm}\hspace*{0.1cm}\hspace*{0.2cm}FILTER: ${S_3[i, j]}$ \\
    \vspace{-0.1cm}
    \caption{\small Initial schedule tree for two matmuls feeding an addition. \normalsize}
    \label{sched:mm-mm-add}
    \vspace{0.2cm}
    \centering
    \lstset{linewidth=\linewidth}
    \begin{lstlisting}[mathescape=true, caption={\small Sum of Matmuls \normalsize},label={lst:mm-mm-add}]
for(i = 0; i $<$ M; ++i)
 for(j = 0; j $<$ N; ++j) {
  for(k = 0; k $<$ K; ++k) {
   C[i, j] = mul_acc(C[i, j], A[i, k], B[k, j]);//S1
   R[i, j] = mul_acc(R[i, j], P[i, k], Q[k, j]);//S2
  }
  Z[i, j] = add(C[i, j], R[i, j]); /*S3*/
 }
    \end{lstlisting}
\end{figure}
\normalsize

Comparing Figures~\ref{fig:sched3-initial} and Figures~\ref{sched:mm-mm-add},
it is clear that the main structural difference between the schedule trees is
that there are three statements involved in the latter and so, three
corresponding filter nodes in sequence. Consequently, the same schedule
transformations can be applied to obtain a schedule tree similar to that in
Figure~\ref{fig:sched3-warp-tile-strip-mined}, except that it would have
three filter nodes -- the first two for the two matmuls and the last one for
the pointwise operations, i.e., at the thread-level all of these operations
are performed in sequence with no interleaving. Memory promotion presents no
additional issues and is handled similarly by avoiding writes of intermediate
results to global memory with only the live-out data being stored to it.

In this example, the input matrices to both the matmuls have the same shape.
However, the overall approach of tiling for blocks with sequence hoisting to
ensure that there is no interleaving of the operations, can be applied even
when the input matrices are of different shapes so long as the matmul outputs
are of the same shape.

\subsection{Pointwise Operations in Prologue}

Consider the scenario where the matmul inputs are results of pointwise
operations as shown in Listing~\ref{lst:mm-relu-prlg}. Since each statement
is tagged with a specification of its expression tree, the compound operation
of ReLU on an input followed by matmul can be represented by a single
statement, with matmul as the root in the expression tree.

\footnotesize
\begin{figure}[h!]
    \centering
    \lstset{linewidth=\linewidth}
    \begin{lstlisting}[mathescape=true, caption={\small ReLU + Matmul \normalsize},label={lst:mm-relu-prlg}]
for(i = 0; i $<$ M; ++i)
 for(j = 0; j $<$ N; ++j)
  for(k = 0; k $<$ K; ++k)
/*S1*/C[i,j] = mul_acc(C[i, j], relu(A[i, k]), B[k,j]);
    \end{lstlisting}
\end{figure}
\normalsize

The initial schedule tree for this loop-nest is structurally similar to that
in Figure~\ref{fig:sched1-initial}. So, all the schedule transformations
discussed in Section~\ref{sec:core1} can be applied. This is followed up with
memory promotion as discussed in Section~\ref{sec:core2} with one crucial
difference. Recall that the global-to-shared memory copy is split into a
global-to-register copy followed by a register-to-shared copy (see
Section~\ref{subsec:global-to-shared}). So, all the data that is reused at
the block level is first moved to registers. Therefore, when we have
pointwise operations in the matmul prologue, instead of moving all this data
to shared memory for reuse, the pointwise operations can be performed on
these data values and the result can instead be stored to shared memory for
reuse. In effect, even though the compound operation in statement $S_1$ could
have multiple input dataspaces, the shared memory footprint remains the same
as that for a matmul without pointwise operations in its prologue. With the
pointwise operations being performed during the data movement, the statement
$S_1$ only needs to perform a macro-MMA using two input fragments and an
accumulator fragment as in case of a single matmul with no prologue. For the
example in Listing~\ref{lst:mm-relu-prlg}, the kernel generated would be the
same as Listing~\ref{lst:matmul-cuda}, except that on lines 20 and 21, the
result obtained by applying a pointwise ReLu on the 8 \textit{fp16} values
which are held in registers, would then be stored to shared memory.
\section{Experimental Evaluation}
\label{sec:experimental}

In this section, we provide an experimental evaluation of the techniques
described in the previous sections 
We implemented these techniques
in the Diesel DSL framework~\cite{elango2018mapl}. Diesel provides a
basic front-end for specifying tensor expressions and to extract
their polyhedral representation. Input to the compiler backend included
various code generation options, namely, block tile sizes, warp tile sizes,
choice of macro-MMA (16$\times$16$\times$8 or 32$\times$32$\times$8).

Automatic tile size selection is beyond the scope of this work. So, for
every benchmark, we generated CUDA kernels using our compiler backend for
various tile size choices and for both the macro-MMA options. All tile sizes
experimented were powers of 2, with 16 being the smallest tile size and
128 being the biggest. The split-K schedule was exercised up to a 4-way split.
The average performance of each of these kernels over 100 runs was obtained
using \textit{nvprof} for 100 problem sizes that were randomly
generated, with problem sizes being multiples of 128 (the largest tile size)
upto a size of 4096. For every problem size, we present the best performance
obtained using our auto-generated kernels.

Baseline versions of the benchmarks were implemented using cuBLAS 10.1 and
cuDNN 7.0. The
\small CUBLAS\textunderscore{GEMM}\textunderscore{DFALT}\textunderscore{TENSOR}\textunderscore{OP} \normalsize
algorithm was chosen to target the tensor cores whereas the pointwise
operations for bias addition and ReLU activation were implemented using the
\textit{cudnnAddTensor} and \textit{cudnnActivationForward} APIs exposed by
cuDNN. The experimental evaluation was performed on an NVIDIA Quadro GV100
GPU with the Volta microarchitecture. The \textit{nvcc} v10.1 compiler was
used to compile all the benchmarks with the options: \small {-O3
-std=c++11
-use\textunderscore{fast}\textunderscore{math} -ccbin g++}
-{arch=compute\textunderscore{70} -code=sm\textunderscore{70}
-expt-relaxed-constexpr.} \normalsize

\begin{figure}[h]
    \centering
    \includegraphics[scale=0.60]{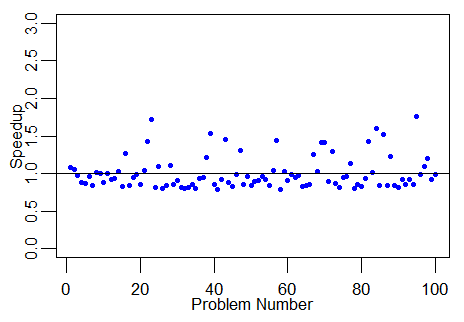}
\caption{\small Performance for {GEMM}.\normalsize}
\label{fig:perf_mma_tn_rand_splitK}
\end{figure}

\paragraph{GEMM}
Figure~\ref{fig:perf_mma_tn_rand_splitK} shows a performance plot of the
speedups obtained using the auto-generated kernels for GEMM over the
\textit{cublasGemmEx} performance. In one in three cases (32\%), the
auto-generated kernels were able to match or outperform the baseline version.
The peak speedup obtained was 1.75$\times$ while it was 0.78$\times$ in the
worst case. For 60 of the 100 problem sizes, the speedup was 0.9$\times$ or
higher and the geometric mean speedup was 0.985$\times$.

\begin{figure}[h]
    \centering
    \includegraphics[scale=0.60]{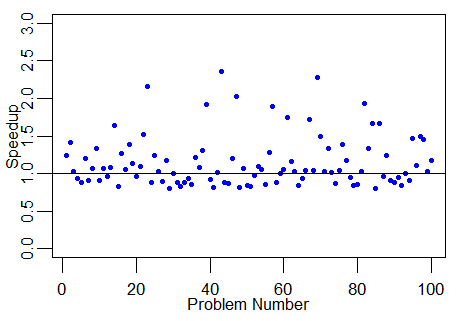}
\caption{\small Performance of prologue fusion: {(ReLU + GEMM)}.\normalsize}
\label{fig:perf_mma_tn_relu_prlg_rand_best_tn_tiles}
\end{figure}

\paragraph{ReLU + GEMM}
In this benchmark, a ReLU pointwise activation function is applied on an
input matrix to GEMM. The auto-generated kernel fuses the ReLU in the GEMM
prologue and the matmul operation into the same device function.
Figure~\ref{fig:perf_mma_tn_relu_prlg_rand_best_tn_tiles} shows a performance
plot of the speedups obtained using the auto-generated kernels for GEMM over
the baseline, which uses \textit{cublasGemmEx} for performing the GEMM and
\textit{cudnnActivationForward}. For 61 problem sizes, the auto-generated
kernels performed at least as well as the baseline version. The peak speedup
obtained was 2.36$\times$ with 0.80$\times$ in the worst case. The geometric
mean speedup was 1.113$\times$. Since the fused kernel is similar to that for
GEMM, except that the ReLU operation is performed on registers during the
data transfer from global to shared memory, the best tile sizes for the fused
kernel correlated with that for GEMM.

\begin{figure}[t]
    \centering
    \includegraphics[scale=0.60]{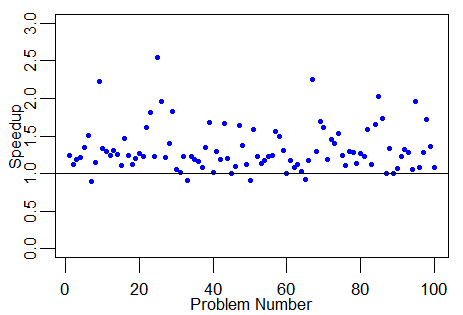}
\caption{\small Epilogue fusion performance: {GEMM} + {Bias} + {Relu}.\normalsize}
\label{fig:perf_mma_tn_bias_relu_rand_splitK}
\end{figure}

\begin{figure}[t]
    \centering
    \includegraphics[scale=0.70]{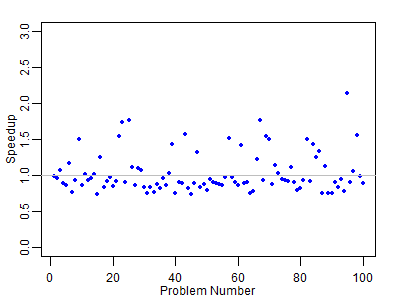}
\caption{\small Kernel fusion performance for {Add(GEMM, GEMM)}.\normalsize}
\label{fig:perf_mma_tn_tn_eplg_rand}
\end{figure}

\paragraph{GEMM + Bias + ReLU}
A performance plot of the speedups obtained using the fused kernels that were
auto-generated for Gemm + Bias + ReLU are shown in
Figure~\ref{fig:perf_mma_tn_bias_relu_rand_splitK}. Compared to
Figure~\ref{fig:perf_mma_tn_rand_splitK}, clearly, kernel fusion moves most
of the speedups over the baseline, with the auto-generated kernels
outperforming the baseline in 94 out of 100 cases. The peak speedup obtained
was 2.55$\times$ with 0.89$\times$ being the worst case speedup, with a mean
speedup of 1.29$\times$.

\paragraph{Add(GEMM, GEMM)}
In this benchmark, the results of two GEMMs of the same shape were fed to an
add operation. Figure~\ref{fig:perf_mma_tn_tn_eplg_rand} shows a plot of the
speedups obtained through a fused kernel over a baseline that used cuBLAS for
gemm and cuDNN for the add operation. For 33 out of the 100 problem sizes,
the fused kernel matches or beats the baseline with 2.13$\times$ peak speedup. The worst case speedup is 0.73$\times$. Overall, with multiple
matmuls being fused into the kernel, we noticed there was greater register
pressure compared to the other fused kernels, necessitating the use of
smaller tiles which while decreasing register pressure were not necessarily
optimal for the matmul computation in the fused kernel.

\section{Related Work}
\label{sec:related-work}

CUDA libraries such as cuBLAS~\cite{cublas} and cuDNN~\cite{chetlur2018cudnn} provide
highly tuned GPU-accelerated implementations of standard basic linear algebra
routines and deep learning primitives. Cutlass~\cite{cutlass} is
a CUDA C++ template library which provides performance that is comparable to
cuBLAS. It features various compute decomposition and data movement strategies
for implementing GEMM, with mixed-precision computation support for 
Volta tensor cores. The tensor core operations in Cutlass are also
implemented using the mma instruction. cuTensor~\cite{cutensor} is a
recent high-performance CUDA library for GPUs with compute
capability greater than or equal to 70. It supports various tensor operations
such as tensor contractions, pointwise operations with support for
pointwise operator fusion.

Polyhedral compilation has been a topic of active research for several
decades~\cite{feautrier92onedim, feautrier92multi}. With a large suite of
tools and libraries~\cite{pluto, pocc, cloog, piplib, loopo}, it has
gradually been incorporated into production compilers such as
RStream~\cite{rstream}, GCC/Graphite~\cite{pop2006graphite},
LLVM/Polly~\cite{polly}.

Domain specific languages such as Polymage~\cite{mullapudi2015asplos} exploit
the sophisticated transformation and code generation capabilities of the
polyhedral framework to automatically generate high performance
implementations of image processing applications. Our work is based on the
Diesel DSL compiler framework developed by Elango et
al~\cite{elango2018mapl}, who also tackled the problem of efficient CUDA
kernel generation for matmul and some epilog fusion scenarios. However, their
focus was more on generating efficient kernels that target traditional CUDA cores. We
build on their work to not only target tensor cores but also cover a wider
range of computation sequences. Other DSL compiler frameworks such as
Halide~\cite{ragan-kelley13pldi}, TVM~\cite{chen2018osdi}, which are
non-polyhedral, separate the notion of the tensor computation from that of
its schedule. Tiramisu~\cite{baghdadi2019tiramisu}, a polyhedral framework to
generate high performance code for GPUs also features a scheduling language
to provide low-level control over the schedule to the user. However, the
schedule primitives for exploiting tensor cores are limited or primarily rely
on the CUDA wmma API for programming them directly~\cite{feng2018}.

Vasilache et al developed Tensor Comprehensions (TC)~\cite{vasilache2020Taco,
vasilache2020Corr}, which leverages the Halide
compiler~\cite{ragan-kelley13pldi} in conjunction with polyhedral compilation
to automatically generate CUDA kernels given a mathematical specification of
a deep learning graph. It uses a modified version of the PPCG compiler
developed by Verdoolaege et al~\cite{verdoolaege2013Taco} with support for operator fusion. While TC handles a larger
class of affine loop-nests, we deal with kernel generation for tensor cores
with a focus on a few common computation idioms. Zerrell et
al~\cite{zerrell2020Corr} developed Stripe, a nested polyhedral intermediate
representation used in the PlaidML~\cite{plaidml} compiler with the facility
to fuse tensor contractions. MLIR~\cite{mlir2019} is an ongoing project which
aims to unify the compiler infrastructure for machine learning by providing
the ability to embed multiple IR dialects in it e.g. linear algebra dialect
or an affine dialect, with a progressive lowering and transformation of IR
dialects. Overall, we believe our work is complementary and could be
integrated with many of these frameworks as a library for targeting tensor
cores.




\vspace{-0.2cm}
\section{Conclusion}
\label{sec:conclusions}

We tackled the problem of automatic generation of efficient CUDA kernels for
computation sequences involving matmul and pointwise operations. To the best
of our knowledge, this is the first work to leverage polyhedral compilation
techniques for exploiting tensor core capabilities on a Volta GPU. In
particular, we relied upon macro-MMA compositions of size
16$\times$16$\times$8 and 32$\times$32$\times$8 implemented using the
\textit{mma.sync.m8n8k4} PTX instruction for targeting tensor cores. Furthermore,
we demonstrated that these techniques can lead to significant 
speedups for a wide range of problem sizes. In the future, we intend to augment
this approach with cost models for automatic tile size selection as well as
generalize it for subsequent GPU micro-architectures such as Turing.


\section*{Acknowledgment}
We thank all contributors to the Diesel compiler -- Venmugil
Elango, Mahesh Ravishankar, Norm Rubin, for creating the framework in which
we could try out the ideas described in this paper. We also thank Bastian
Hagedorn for his comments on leveraging the mma instructions for targeting
Volta tensor cores.

\bibliographystyle{IEEEtran}
\bibliography{bib}

\begin{thebibliography}{10}
\providecommand{\url}[1]{#1}
\csname url@samestyle\endcsname
\providecommand{\newblock}{\relax}
\providecommand{\bibinfo}[2]{#2}
\providecommand{\BIBentrySTDinterwordspacing}{\spaceskip=0pt\relax}
\providecommand{\BIBentryALTinterwordstretchfactor}{4}
\providecommand{\BIBentryALTinterwordspacing}{\spaceskip=\fontdimen2\font plus
\BIBentryALTinterwordstretchfactor\fontdimen3\font minus
  \fontdimen4\font\relax}
\providecommand{\BIBforeignlanguage}[2]{{%
\expandafter\ifx\csname l@#1\endcsname\relax
\typeout{** WARNING: IEEEtran.bst: No hyphenation pattern has been}%
\typeout{** loaded for the language `#1'. Using the pattern for}%
\typeout{** the default language instead.}%
\else
\language=\csname l@#1\endcsname
\fi
#2}}
\providecommand{\BIBdecl}{\relax}
\BIBdecl

\bibitem{cublas}
NVIDIA, ``cublas,'' 2019, \goodbreak
  https://docs.nvidia.com/cuda/cublas/index.html.

\bibitem{cutlass}
------, ``Cuda templates for linear algebra subroutines,'' 2019, \goodbreak
  https://github.com/NVIDIA/cutlass.

\bibitem{chetlur2018cudnn}
\BIBentryALTinterwordspacing
S.~Chetlur, C.~Woolley, P.~Vandermersch, J.~Cohen, J.~Tran, B.~Catanzaro, and
  E.~Shelhamer, ``cudnn: Efficient primitives for deep learning,'' \emph{CoRR},
  vol. abs/1410.0759, 2014. [Online]. Available:
  \url{http://arxiv.org/abs/1410.0759}
\BIBentrySTDinterwordspacing

\bibitem{abadi2016osdi}
\BIBentryALTinterwordspacing
M.~Abadi, P.~Barham, J.~Chen, Z.~Chen, A.~Davis, J.~Dean, M.~Devin,
  S.~Ghemawat, G.~Irving, M.~Isard, M.~Kudlur, J.~Levenberg, R.~Monga,
  S.~Moore, D.~G. Murray, B.~Steiner, P.~A. Tucker, V.~Vasudevan, P.~Warden,
  M.~Wicke, Y.~Yu, and X.~Zheng, ``Tensorflow: {A} system for large-scale
  machine learning,'' in \emph{12th {USENIX} Symposium on Operating Systems
  Design and Implementation, {OSDI} 2016, Savannah, GA, USA, November 2-4,
  2016}, 2016, pp. 265--283. [Online]. Available:
  \url{https://www.usenix.org/conference/osdi16/technical-sessions/presentation/abadi}
\BIBentrySTDinterwordspacing

\bibitem{paszke2019nips}
\BIBentryALTinterwordspacing
A.~Paszke, S.~Gross, F.~Massa, A.~Lerer, J.~Bradbury, G.~Chanan, T.~Killeen,
  Z.~Lin, N.~Gimelshein, L.~Antiga, A.~Desmaison, A.~K{\"{o}}pf, E.~Yang,
  Z.~DeVito, M.~Raison, A.~Tejani, S.~Chilamkurthy, B.~Steiner, L.~Fang,
  J.~Bai, and S.~Chintala, ``Pytorch: An imperative style, high-performance
  deep learning library,'' in \emph{Advances in Neural Information Processing
  Systems 32: Annual Conference on Neural Information Processing Systems 2019,
  NeurIPS 2019, 8-14 December 2019, Vancouver, BC, Canada}, 2019, pp.
  8024--8035. [Online]. Available:
  \url{http://papers.nips.cc/paper/9015-pytorch-an-imperative-style-high-performance-deep-learning-library}
\BIBentrySTDinterwordspacing

\bibitem{cudaToolkit}
NVIDIA, ``Cuda toolkit documentation,'' 2019, \goodbreak
  https://docs.nvidia.com/cuda/parallel-thread-execution/index.html\#warp-level-matrix-fragment-mma-884.

\bibitem{wmma2017}
------, ``Programming tensor cores in cuda 9,'' 2017, \goodbreak
  https://devblogs.nvidia.com/programming-tensor-cores-cuda-9/.

\bibitem{uday08pldi}
U.~Bondhugula, A.~Hartono, J.~Ramanujam, and P.~Sadayappan, ``A practical
  automatic polyhedral program optimization system,'' in \emph{PLDI}, Jun 2008.

\bibitem{verdoolaege2013Taco}
\BIBentryALTinterwordspacing
S.~Verdoolaege, J.~C. Juega, A.~Cohen, J.~I. G{\'{o}}mez, C.~Tenllado, and
  F.~Catthoor, ``Polyhedral parallel code generation for {CUDA},''
  \emph{{TACO}}, vol.~9, no.~4, pp. 54:1--54:23, 2013. [Online]. Available:
  \url{https://doi.org/10.1145/2400682.2400713}
\BIBentrySTDinterwordspacing

\bibitem{baskaran08ics}
M.~Baskaran, U.~Bondhugula, S.~Krishnamoorthy, J.~Ramanujam, A.~Rountev, and
  P.~Sadayappan, ``{A Compiler Framework for Optimization of Affine Loop Nests
  for GPGPUs},'' in \emph{{ACM Intl. conference on Supercomputing (ICS)}}, Jun.
  2008.

\bibitem{vasilache2018tcCoRR}
\BIBentryALTinterwordspacing
N.~Vasilache, O.~Zinenko, T.~Theodoridis, P.~Goyal, Z.~DeVito, W.~S. Moses,
  S.~Verdoolaege, A.~Adams, and A.~Cohen, ``Tensor comprehensions:
  Framework-agnostic high-performance machine learning abstractions,''
  \emph{CoRR}, vol. abs/1802.04730, 2018. [Online]. Available:
  \url{http://arxiv.org/abs/1802.04730}
\BIBentrySTDinterwordspacing

\bibitem{mlir2019}
\BIBentryALTinterwordspacing
C.~Lattner, J.~A. Pienaar, M.~Amini, U.~Bondhugula, R.~Riddle, A.~Cohen,
  T.~Shpeisman, A.~Davis, N.~Vasilache, and O.~Zinenko, ``{MLIR:} {A} compiler
  infrastructure for the end of moore's law,'' 2020. [Online]. Available:
  \url{https://arxiv.org/abs/2002.11054}
\BIBentrySTDinterwordspacing

\bibitem{schreiber2020impact}
S.~V. M. K.~R. Schreiber and H.~Kamepalli, ``Generating simd instructions for
  cerebras cs-1 using polyhedral compilation techniques,'' 2020.

\bibitem{isl}
\BIBentryALTinterwordspacing
S.~Verdoolaege, ``\emph{isl}: An integer set library for the polyhedral
  model,'' in \emph{Mathematical Software - {ICMS} 2010, Third International
  Congress on Mathematical Software, Kobe, Japan, September 13-17, 2010.
  Proceedings}, 2010, pp. 299--302. [Online]. Available:
  \url{https://doi.org/10.1007/978-3-642-15582-6\_49}
\BIBentrySTDinterwordspacing

\bibitem{verdoolaege14IMPACT}
S.~Verdoolaege, S.~Guelton, T.~Grosser, and A.~Cohen, ``Schedule trees,'' in
  \emph{IMPACT}, 01 2014.

\bibitem{ragan-kelley13pldi}
J.~Ragan-Kelley, C.~Barnes, A.~Adams, S.~Paris, F.~Durand, and S.~P.
  Amarasinghe, ``Halide: a language and compiler for optimizing parallelism,
  locality, and recomputation in image processing pipelines,'' in \emph{ACM
  SIGPLAN symposium on Programming Languages Design and Implementation}, 2013,
  pp. 519--530.

\bibitem{elango2018mapl}
\BIBentryALTinterwordspacing
V.~Elango, N.~Rubin, M.~Ravishankar, H.~Sandanagobalane, and V.~Grover,
  ``Diesel: {DSL} for linear algebra and neural net computations on gpus,'' in
  \emph{Proceedings of the 2nd {ACM} {SIGPLAN} International Workshop on
  Machine Learning and Programming Languages, MAPL@PLDI 2018, Philadelphia, PA,
  USA, June 18-22, 2018}, 2018, pp. 42--51. [Online]. Available:
  \url{https://doi.org/10.1145/3211346.3211354}
\BIBentrySTDinterwordspacing

\bibitem{cutensor}
NVIDIA, ``cutensor: A high-performance cuda library for tensor primitives,''
  2019, \goodbreak https://docs.nvidia.com/cuda/cutensor/index.html.

\bibitem{feautrier92onedim}
P.~Feautrier, ``Some efficient solutions to the affine scheduling problem: Part
  {I}, one-dimensional time,'' \emph{Intl. Journal of Parallel Programming},
  vol.~21, no.~5, pp. 313--348, 1992.

\bibitem{feautrier92multi}
------, ``Some efficient solutions to the affine scheduling problem: Part {II},
  multidimensional time,'' \emph{Intl. Journal of Parallel Programming},
  vol.~21, no.~6, pp. 389--420, 1992.

\bibitem{pluto}
``{PLUTO: An automatic polyhedral parallelizer and locality optimizer for
  multicores},'' http://pluto-compiler.sourceforge.net.

\bibitem{pocc}
``{POCC: Polyhedral compiler collection},'' http://pocc.sourceforge.net.

\bibitem{cloog}
``{CLooG: The Chunky Loop Generator},'' http://www.cloog.org.

\bibitem{piplib}
``{PIP: The Parametric Integer Programming Library},'' http://www.piplib.org.

\bibitem{loopo}
``{The LooPo Project - Loop parallelization in the polytope model},''
  http://www.fmi.uni-passau.de/loopo.

\bibitem{rstream}
``{RSTREAM - High Level Compiler, Reservoir Labs},'' http://www.reservoir.com.

\bibitem{pop2006graphite}
S.~Pop, A.~Cohen, C.~Bastoul, S.~Girbal, G.~Silber, and N.~Vasilache,
  ``Graphite: Loop optimizations based on the polyhedral model for gcc,'' 2006.

\bibitem{polly}
T.~Grosser, H.~Zheng, R.~Aloor, A.~Simbürger, A.~Großlinger, and L.-N.
  Pouchet, ``Polly: Polyhedral optimization in {LLVM},'' in \emph{IMPACT},
  2011.

\bibitem{mullapudi2015asplos}
R.~T. Mullapudi, V.~Vasista, and U.~Bondhugula, ``Polymage: Automatic
  optimization for image processing pipelines,'' in \emph{Intl. Conference on
  Architectural Support for Programming Languages and Operating Systems}, ser.
  ASPLOS '15, 2015, pp. 429--443.

\bibitem{chen2018osdi}
\BIBentryALTinterwordspacing
T.~Chen, T.~Moreau, Z.~Jiang, L.~Zheng, E.~Q. Yan, H.~Shen, M.~Cowan, L.~Wang,
  Y.~Hu, L.~Ceze, C.~Guestrin, and A.~Krishnamurthy, ``{TVM:} an automated
  end-to-end optimizing compiler for deep learning,'' in \emph{13th {USENIX}
  Symposium on Operating Systems Design and Implementation, {OSDI} 2018,
  Carlsbad, CA, USA, October 8-10, 2018}, 2018, pp. 578--594. [Online].
  Available: \url{https://www.usenix.org/conference/osdi18/presentation/chen}
\BIBentrySTDinterwordspacing

\bibitem{baghdadi2019tiramisu}
\BIBentryALTinterwordspacing
R.~Baghdadi, J.~Ray, M.~B. Romdhane, E.~D. Sozzo, A.~Akkas, Y.~Zhang,
  P.~Suriana, S.~Kamil, and S.~P. Amarasinghe, ``Tiramisu: {A} polyhedral
  compiler for expressing fast and portable code,'' in \emph{{IEEE/ACM}
  International Symposium on Code Generation and Optimization, {CGO} 2019,
  Washington, DC, USA, February 16-20, 2019}, 2019, pp. 193--205. [Online].
  Available: \url{https://doi.org/10.1109/CGO.2019.8661197}
\BIBentrySTDinterwordspacing

\bibitem{feng2018}
``How to optimize convolution using tensorcores,'' 2018, \goodbreak
  https://docs.tvm.ai/tutorials/optimize/opt\_conv\_tensorcore.html.

\bibitem{vasilache2020Taco}
\BIBentryALTinterwordspacing
N.~Vasilache, O.~Zinenko, T.~Theodoridis, P.~Goyal, Z.~DeVito, W.~S. Moses,
  S.~Verdoolaege, A.~Adams, and A.~Cohen, ``The next 700 accelerated layers:
  From mathematical expressions of network computation graphs to accelerated
  {GPU} kernels, automatically,'' \emph{{TACO}}, vol.~16, no.~4, pp.
  38:1--38:26, 2020. [Online]. Available: \url{https://doi.org/10.1145/3355606}
\BIBentrySTDinterwordspacing

\bibitem{vasilache2020Corr}
\BIBentryALTinterwordspacing
------, ``Tensor comprehensions: Framework-agnostic high-performance machine
  learning abstractions,'' \emph{CoRR}, vol. abs/1802.04730, 2018. [Online].
  Available: \url{http://arxiv.org/abs/1802.04730}
\BIBentrySTDinterwordspacing

\bibitem{zerrell2020Corr}
\BIBentryALTinterwordspacing
T.~Zerrell and J.~Bruestle, ``Stripe: Tensor compilation via the nested
  polyhedral model,'' \emph{CoRR}, vol. abs/1903.06498, 2019. [Online].
  Available: \url{http://arxiv.org/abs/1903.06498}
\BIBentrySTDinterwordspacing

\bibitem{plaidml}
Intel, ``Plaidml,'' 2019, \goodbreak https://www.intel.ai/plaidml.

\end{thebibliography}


\end{document}